\RequirePackage{fix-cm}
\documentclass[smallextended]{svjour3}       
\smartqed  
\usepackage{graphicx}
\usepackage{mathtools}
\usepackage{natbib}
\usepackage{hyperref}
\newcommand{\alphaox}{\alpha_{\rm OX}}
\newcommand{\fuv}{F_{\rm UV}}
\newcommand{\luv}{L_{\rm UV}}
\newcommand{\fx}{F_{\rm X}}
\newcommand{\Lx}{L_{\rm X}}
\newcommand{\ho}{$H_{0}$}
 
\newcommand{\sne}{SNe~Ia}
\newcommand{\sn}{SN~Ia}

\usepackage[dvipsnames]{xcolor}
\usepackage{amsmath} 
\usepackage{amssymb}

\journalname{Space Science Review}
\begin{document}

\title{Astronomical Distance Determination in the Space Age 
}
\subtitle{Secondary distance indicators}

\titlerunning{Secondary distance indicators}        

\author{Bo\. zena Czerny         \and
	Rachael Beaton \and
        Micha\l~ Bejger  \and
        Edward Cackett \and
       Massimo Dall'Ora \and
       R. F. L.  Holanda \and
       Joseph B. Jensen        \and
        Saurabh W. Jha \and 
        Elisabeta Lusso \and
         Takeo Minezaki \and
        Guido Risaliti \and
        Maurizio Salaris   \and
        Silvia Toonen \and
        Yuzuru Yoshii \and
       }
\authorrunning{B. Czerny et al.} 

\institute{B. Czerny \at
              Center for Theoretical Physics, Polish Academy of Sciences, Al. Lotnikow 32/46, 02-668 Warsaw, Poland\\
              \email{bcz@cft.edu.pl}           
           \and
           R. Beaton  \at
              The Observatories of the Carnegie Institution of Washington, 813 Santa Barbara St., Pasadena, CA 91101, U
              \and
           M. Bejger \at
       Copernicus Astronomical Center, Polish Academy of Sciences, Bartycka 18, 00-716 Warsaw, Poland
       \and 
         E. Cackett \at
            Department of Physics \& Astronomy, Wayne State University, 666 W. Hancock St, Detroit, MI 48201, US
            \and
              M. Dall'Ora \at
    INAF, Osservatorio Astronomico di Capodimonte, via Moiariello 16, 80131 Napoli, Italy
              \and
             R. F. L.  Holanda \at
            Departamento de F\'{\i}sica, Universidade Federal de Sergipe, 49100-000, Aracaju - SE, Brazil
              \email{holanda@uepb.edu.br}                      
             \and
            J. B. Jensen \at
              Utah Valley University \\
              \email{jjensen@uvu.edu}
              \and
          S. W. Jha \at
          Department of Physics and Astronomy, Rutgers, The State University of New Jersey, Piscataway, NJ 08854, USA
          \and
          E. Lusso \at 
          INAF - Arcetri Astrophysical Observatory, Largo E. Fermi 5, I-50125 Firenze, Italy \\
          Centre for Extragalactic Astronomy, Department of Physics, Durham University, South Road, Durham, DH1 3LE, UK\\
          \and
          T. Minezaki \at
          Institute of Astronomy, Graduate School of Science, The University of Tokyo, 2-21-1 Osawa, Mitaka, Tokyo 181-0015, Japan, \email{minezaki@ioa.s.u-tokyo.ac.jp}
          \and
          G. Risaliti \at
          INAF - Arcetri Astrophysical Observatory, Largo E. Fermi 5, I-50125 Firenze, Italy
          \and
           M. Salaris \at
              Astrophysics Research Institute, Liverpool John Moores University, IC2, Liverpool Science Park, 146 Brownlow Hill, Liverpool L3 5RF, UK \email{M.Salaris@ljmu.ac.u}
            \and
          S. Toonen \at
              Anton Pannekoek Institute for Astronomy, University of Amsterdam, 1090 GE Amsterdam, The Netherlands
            \and
         Y. Yoshii \at
       Institute of Astronomy, Graduate School of Science, The University of Tokyo,
       2-21-1 Osawa, Mitaka, Tokyo 181-0015, Japan,\\
       Steward Observatory, University of Arizona,
       933 North Cherry Avenue, Room N204,\\ Tucson, AZ 85721-0065, USA,\\
       \email{yoshii@ioa.s.u-tokyo.ac.jp}\\
}

\date{Received: date / Accepted: date}

\maketitle

\begin{abstract}
The formal division of the distance indicators into primary and secondary leads to difficulties in description of methods which can actually be used in two ways: with, and without the support of the other methods for scaling. Thus instead of concentrating on the scaling requirement we concentrate on all methods of distance determination to extragalactic sources which are designated, at least formally, to use for individual sources. Among those, the Supernovae Ia is clearly the leader due to its enormous success in determination of the expansion rate of the Universe. However, new methods are rapidly developing, and there is also a progress in more traditional methods. We give a general overview of the methods but we mostly concentrate on the most recent developments in each field, and future expectations.
\end{abstract}

\section{Introduction}
\label{intro}

Measurement of the distance to extragalactic objects is an important task. For objects at large distances the measurement of the redshift was frequently considered as the measurement of the distance but precision cosmology requires the knowledge of both the distance and the redshift.

The measurement of the distances to extragalactic objects was for years based on a carefully set subsequent steps of the cosmic ladder: direct distance measurement methods allowed to formulate scaling relations which then could be applied to more distant objects, and with a few such steps cosmological distances were achieved. The fundamental direct method of measuring the distance in astronomy is to measure the trigonometric parallax of a given source, reflecting the motion of the Earth around the Sun. Hipparcos mission measured parallaxes up to a few hundred parsecs, Hubble Space Telescope reached up to 5 kpc in dedicated observations. Now \emph{Gaia} mission is collecting the data.
It will cover with precisions of 10\% a volume of radius ~10 kpc, which encompasses $<$ 1\% of the stars in our Galaxy by number, but by volume is very large and will cover a large fraction of the Milky Way with reasonable accuracy. 
Parallaxes of the Cepheid stars, combined with measurements in nearby galaxies, calibrate the Leavitt Law for Cepheids, and Cepheids are accessible to ~30 Mpc with the Hubble Space Telescope.
Further out, \sne~ calibrated with the help of the Cepheids take over, covering cosmological distances up to the redshift $\sim 2$. This method replaced older methods of covering most distant part of the ladder, like Tully-Fisher relation.

Modern astronomy still strongly relies on indirect measurements which in turn rely in some intermediate steps but in the aim to meet expectations of the precision cosmology the intermediate steps are refined, new shortcut developed and some of the basically indirect methods might be potentially used as direct methods under some physically well motivated assumptions. In the Local Group the use of the eclipsing binary stars allows to skip the step of the parallax measurement. A very important direct distance measurement is possible for the spatially resolved water maser in the galaxy NGC 4258. The impressive jump directly to very large distances opened with observations of gravitational lensing of quasars and the measurement of the time delays between the quasar images which allows their use as a ruler. Another direct method can be applied to galaxy clusters. This method uses the Sunyeav-Zeldovich effect. Extremely important direct methods are the Barion Acoustic Oscillation method and the observations of the fluctuations of the Cosmic Microwave Background. In both cases the (statistical) analysis of the sizes of the fluctuations allows to get directly the cosmological constraints. Another statistical method to obtain the cosmological constraints is gravitational lensing. Now, with firm detection of gravitational waves, the possibility of the direct determination of the distance to the gravitational wave source from the shape of its signal also gains importance. All these methods are described below.

We do not follow the strict definition of indirect method but in this section we concentrate on {\it predominantly} indirect methods which can be applied to {\it single specific extragalactic objects} like supernovae, galaxies, galaxy clusters, active galaxies or gamma-ray bursts. Global statistical methods, including the results based on the Cosmic Microwave Background, will be discussed in a separate Chapter. We also do not include in this Chapter strong gravitational lensing since it will be more natural to address this issue in the Chapter where weak lensing is discussed. 

We outline the methods shortly, with references to the basic books or reviews, and we concentrate on most recent developments in the field. Whenever possible, we discuss the issue of the method cross-calibrations. 

The order of the methods' presentation is roughly adjusted to their past and current popularity and role in astronomy, but not necessarily reflects their potential for development in the near future.  

\section{Supernovae Ia}
\label{sec:SN}


Type Ia supernovae (SNe~Ia) are defined by the lack of hydrogen and helium lines around the maximum light 
and the presence of strong SiII absorption features. Their peak absolute magnitudes are bright 
($M_B$ typically between $-$18.5 and $-19.5$~mag) and moreover for the majority of them \citep[about 64\% of all 
SN~Ia events, denoted as \lq{normal}\rq, see][]{li} 
the light curves are remarkably similar, in the sense 
that they display a well defined relation between the peak absolute brightness and the width of their lightcurve 
\citep{phillips_1993}. This characteristic makes SN~Ia events very good cosmological distance indicators, and led to the discovery 
of the accelerated expansion of the universe \citep{riess, perlm}. 

\sne~ are intrinsically luminous \citep{riess_2016},  thus, even with modest aperture telescopes \sne~ can be discovered over large volumes; indeed, the All Sky Survey for Supernova \citep[ASAS-SN;][]{shappee_2014} uses telescopes with only a 14 cm aperture to discover a total of 242 confirmed SNe, of which 183 were \sn~ (76\%) over its first 18 months of dual-hemisphere operations (01 May 2014 to 31 December 2015) alone within $m_V$~\textless~17 \citep[][]{asassn_15}\footnote{See also \citet{asassn_1314} and \citet{asassn_16} for the 2013-2014 and 2016 statistics, respectively}.
Their intrinsic luminosity also permits their discovery at significant redshift; the previous most distant \sn~  SN\,UDS10Wil at $z$=1.914, which was discovered in the Cosmic Assembly Near-infrared Deep Extragalactic Legacy Survey \citep[CANDELS, PI: Faber \& Ferguson; see][for details]{jones_2013}, was recently replaced with another 15 in the range from 1.9$ <$ z $< $2.3 presented in \citet{riess2017}. \sne~ in the redshift range from 0.2 \textless z \textless 0.8 are now routinely discovered and used for cosmological purposes; in the first two years of the ESSENCE project, 52 \sne~ were discovered in this redshift range \citep{metheson_2005}. 
 
\sne, however, are not trivial objects to study and understand. 
Paramount among the difficulties of \sne~ is that their progenitors and the exact physics of their ignition mechanisms are actively debated (see Sec.~\ref{sec:SN_progenitors}).
Despite this, empirical measurements of the \sne, as a class, show remarkable homogeneity and, due to their frequency and ubiquity, \sne~ are likely to remain our tools for an en masse probe of the distant Universe. 
Figure \ref{fig:hubbleflow} is a Hubble Diagram, a plot of distance versus redshift, for \sne~ within $z$~\textless~0.1 that demonstrates the remarkable precision ($\sigma_{\sne}$~$\sim$~0.12 mag or 6\% in distance) of \sne~ as a class. 
With a sample size of \textgreater~200 (and growing), this stage of the extra-galactic distance scale is, by far, the most well constrained and provides a random uncertainty of $\sim$0.7\% to the determination of the local expansion rate of the Universe (the Hubble constant or $H_{0}$) \citep[for details see discussion in][]{beaton_2016}.   
In this Section, the key observational components and considerations required to construct the dataset in Figure \ref{fig:hubbleflow} are described. 

\subsection{Early History of SNe}

\citet{wood-vasey_2007} present a short contextual history of the role of SNe for cosmology that we will expand upon herein. 
Within the context of the `Great Debate', \citet{shapley_1919}  argued that intrinsic brightness of SN\,1885A found within the Andromeda galaxy of $M$ = -16 mag was ``out of the question'' if Andromeda was, indeed, an `extragalactic nebulae.' 
Thus, Shapley argued, the island universe hypothesis for the nature of the nebulae was likely incorrect owing to this extreme luminosity. 
In his landmark paper describing the Cepheid variables discovered in Andromeda between 1923-1928, \citep{hubble_1929} commented on the existence of ``a mysterious class of exceptional novae which attain luminosities that are respectable fractions of the total luminosities of the systems in which they appear.'' 
Since Hubble's work provided strong evidence that the nebulae were, indeed, extra-galactic in nature \citep{code_1999}, this new class of variables was later classified as `super-novae' by \citet{baade_1934} because the approximate amount of energy released by these objects over 25 days ``is equivalent to 10$^{7}$ years of solar radiation of the present strength.''

\citet{minkowski_1941} was the first to apply a classification scheme to the supernovae, which was based on the presence or absence of hydrogen emission lines in the spectra (the type II and type I, designations respectively).
As Minkowski notes, the spectra for SNe of type I seemed remarkably homogeneous in support of the postulation by \citet{wilson_1939} that such objects could be employed for cosmological exploration.
Indeed, the discovery and utilization of SNe for distances relies on their photometric  properties, but the classification of SNe into phenomenological subtypes is largely reliant on their spectroscopy.
The dual-need of spectroscopic {\emph {and}} photometric classification is a challenge for the accumulations of the anticipated large samples to be discovered in the 2020's by large scale transient monitoring programs like the Large Synoptic Survey Telescope  \citep[LSST; see discussions in][]{lsst_spec,lsst_sci_book}.

Moreover, with the advent of long-term spectroscopic and photometric monitoring of a large number of \sne, there are now additional sub-classifications based on both spectroscopic and photometric diagnostics \citep[see discussion in][]{parrent_2014}. 

While both our theoretical understanding and experimental census of \sne~ has increased dramatically, many of the original mysteries regarding the \sne~ posed by \citet{baade_1934}, when their class was initially defined remain, as vibrant areas of research nearly a century later.

\subsection{\sn~ as Standardize-able Candles}
While their early homogeneity was suggestive of their utility as standard candles as early as \citet{wilson_1939}, the ability to be used as such occurred only in the last 30 years. \citet{phillips_1993}, expanding on earlier work by \citet{Pskovskii_1977}, reported on an apparent empirical correlation between the absolute luminosity of a \sn~ and its the rate-of-decline in optical lightcurves ($BVI$). 
The decline rate is the change in brightness of a given band over the 15 days post maximum and is designated $\Delta m_{15}(\lambda)$; a typical decline rate is $\Delta m_{15}(\lambda)$=1, but rates can vary between 0.9~\textless~$\Delta m_{15}(\lambda)$~\textless 1.8 \citep{phillips_2012}.
The $\Delta m_{15}$ parameter is relatively straight forward to measure with appropriate light curves and acts as a scale factor through which light curves of \sn~ are standardized and used as distance indicators.
In works by \citet{hamuy_1995} and \citet{riess_1996}, it was clear that \sne~ corrected for their decline rate provided precise distances. 

\begin{figure*}
\centering
\includegraphics[width=1.0\textwidth]{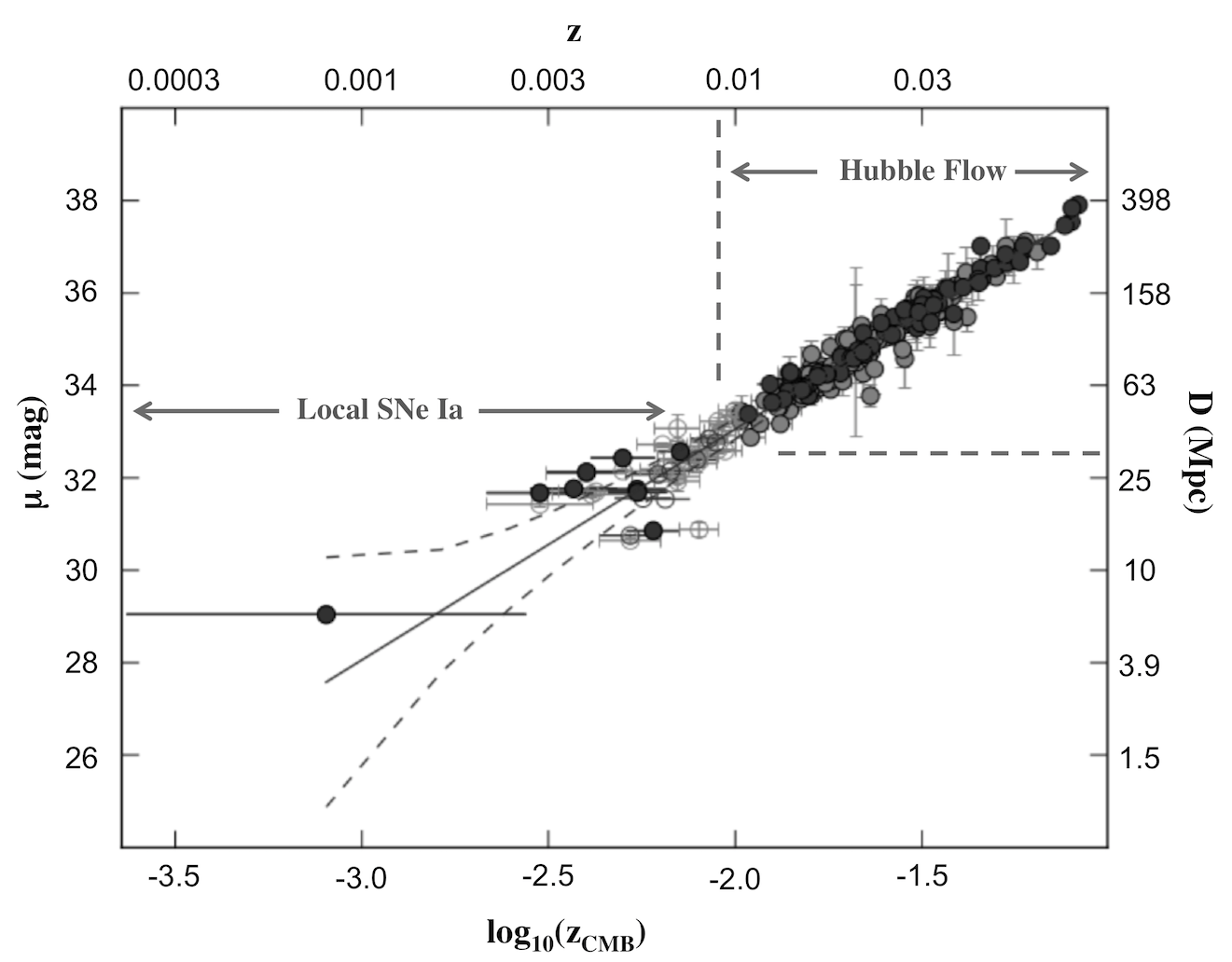}
\caption{ \label{fig:hubbleflow}
Hubble diagram for \sn~ in the \emph{Carnegie Supernova Project} (dark grey) and the CfA4 survey (grey). 
The scatter for the \sn~ in the `Hubble Flow' sample ($z$ \textgreater 0.01) is 0.12 mag or 6\% in distance, which when averaged over the 200 \sn~ in the sample, results in a total uncertainty of 0.7\% \citep[][]{beaton_2016}. The effect of peculiar motions on the `local' sample are evident by their large scatter.
The `local' sample is far less complete, with only eight galaxies in this visualization; \citet{riess_2016} brought another 11 \sn~ into the `local' sample. [Adapted from \citet{beaton_2016}].}
\end{figure*}

The physics underlying this relationship is that intrinsically brighter \sne~ take longer to fade than intrinsically fainter \sne~ and this is largely thought to be due to different amounts of $^{56}$Ni contributing to the early time light curve, though the reason for this could be due to differences in either the total mass or its distribution in the explosion \citep[see][for detailed discussions]{kasen_2007,parrent_2014}. 
Recent energy has been focused on the near-infrared properties of \sne, which show far more homogeneity than the optical and are less sensitive to dust \citep[a detailed review is given by][]{phillips_2012}.
Indeed, in Figure \ref{fig:hubbleflow}, the dark grey points from the Carnegie Supernova Project (CSP) show a tighter dispersion than those in grey from Harvard-Smithsonian Center for Astrophysics (CfA4), precisely because of the CSP use of near-infrared photometry.

Since the large scale demonstration of \sne~ as standardizable candles, they have been studied at a number of regimes; these are:
(i) local ($z$~\textless~0.03), 
(ii) intermediate (0.03~\textless~$z$~\textless~0.3), 
(iii) distant ($z$~\textgreater~0.3), 
and (iv) high-z ($z$~\textgreater~1.2). 
The \sn~ discovered within each regime are used for different goals.
Local \sne~ can simultaneously constrain progenitor properties and serve as calibration sources.
Intermediate \sne~ are safely in the Hubble Flow and used to measure \ho.
Distant and high-z samples are used to trace the acceleration of the Universe ($w$). 
Thus, large collaborations have been formed to find, study, and use \sne~ and these generally separate into redshift (and brightness) regimes.

Amongst numerous efforts in the community, several of note are as follows: 
 the CfA survey \citep{riess_1999,jha_2006,hicken_2009},
 the Carnegie Supernova Project \citep[CSP][]{hamuy_2006,freedman_2009,folatelli_2010}, 
 ESSENCE \citep{miknaitis_2007,wood-vasey_2007,foley_2008,foley_2009},
 SH0ES \citep{riess_2004a,riess_2004b,riess_2007,riess_2009}, 
 the Supernova Legacy Survey \citep[SNLS][and references therein]{sullivan_2011},
 and the Foundation Survey within Pan-STARRS \citep[e.g,][]{foley_2017}. 

In addition to these largely ground based efforts, there are devoted space-based programs, one of note being the SWIFT Supernova Program that is capable of producing early- to late-time UV measurements useful for understanding the local \sn~ environs \citep{brown_2015a}.

A database of SNe discoveries (of all types), largely driven by transient monitoring programs, is maintained on the \emph{Latest Supernovae} website \citep{galyam_2013}\footnote{Available: \url{http://www.rochesterastronomy.org/snimages/}},
 which includes discovery imaging and classifications on a day-to-day basis. 
Individual teams maintain their own databases (both publicly available and proprietary), 
 but the \emph{Open Supernova Catalog}\footnote{Available: \url{https://sne.space/}} provides a uniform access point for both original observations and derived parameters with extensive citations \citep{opensupernovacatalog}. 
 
\subsection{\sn~ as Cosmological Probes}

The role of \sne~ in the construction of our concordance cosmology is hard to understate.
They provide the largest sample of tracers for measuring the Hubble constant (as demonstrated in Figure \ref{fig:hubbleflow}) and their intrinsic luminosity gives us access to the evolution of the Hubble constant over time. 
The latter led to the discovery of the accelerating Universe \citep{perlmutter_1997,perlmutter_1998,perlmutter_1999,riess_1998,schmidt_1998}.
Of particular interest are the proceedings of the Centennial Symposium for the Carnegie Institution of Washington, entitled ``Measuring and Modeling the Universe,'' which carefully reflects on the establishment of the concordance cosmology \citep{freedman_2004}.
More recent reviews for which \sne~ play a central role are: the Hubble constant by \citet{freedman_2010}, for Dark Energy and the accelerating Universe by \citet{frieman_2008}, and a prospectus for the long term impact on physics by \citet{goobar_2011}.
The hundreds of citations to these review articles indicate the speed at which these fields are evolving. 

Indeed, in a span of less than 20 years the `factor of two controversy' in the value of the Hubble constant, ultimately resolved with the {\emph HST} Key Project \citep{freedman_2001}, has been recast as a 3-$\sigma$ discrepancy between the values measured from \sne~  \citep[e.g., by calibration of Figure \ref{fig:hubbleflow};][]{riess_2016,freedman_2012} and that inferred from modeling anisotropies in the cosmic microwave background \citep[most recently,][]{planck_2016}. 

As is demonstrated by \citet[][their Fig.~1]{beaton_2016}, as the uncertainties in either measurement technique have progressively decreased, the discrepancy between the two techniques has only grown. 
As described in \citet{freedman_2017}, the \sn~ derived cosmology `is at a crossroads.'
Thus, it is meaningful to understand the observational metrics through which we understand and characterize the \sn, which are driven by the wealth of data that can be collected in the near-field.

For the purposes of the distance scale, \sne~ in the local and intermediate redshift regime are of the greatest interest, with measurements made on those samples ultimately informing their use at ever more distant regimes.
Those samples used for cosmology (z \textgreater 0.3) ultimately drive a set of criteria that determine the ``suitability'' of a local or intermediate \sn~ for direct calibration. 
These restrictions are motivated by the desire to use the same intrinsic distance indicator in all regimes and, due to observability constraints, many of the fundamental parameters (such as the dispersion within a class of \sn~ sample) are only well-characterized in local or intermediate samples (to be discussed in a later section). 
The requirements from \citet{riess_2016} are:
 (i) observations taken on modern CCD detectors, (ii) not highly reddenned ($A_{V}$ \textless 0.5), (iii) discovered before peak brightness, (iv) have `typical' spectroscopic designations, and (v) relatively homogeneous photometric data (same filters, etc.).
 

\begin{figure*}

\centering

\includegraphics[width=1.0\textwidth]{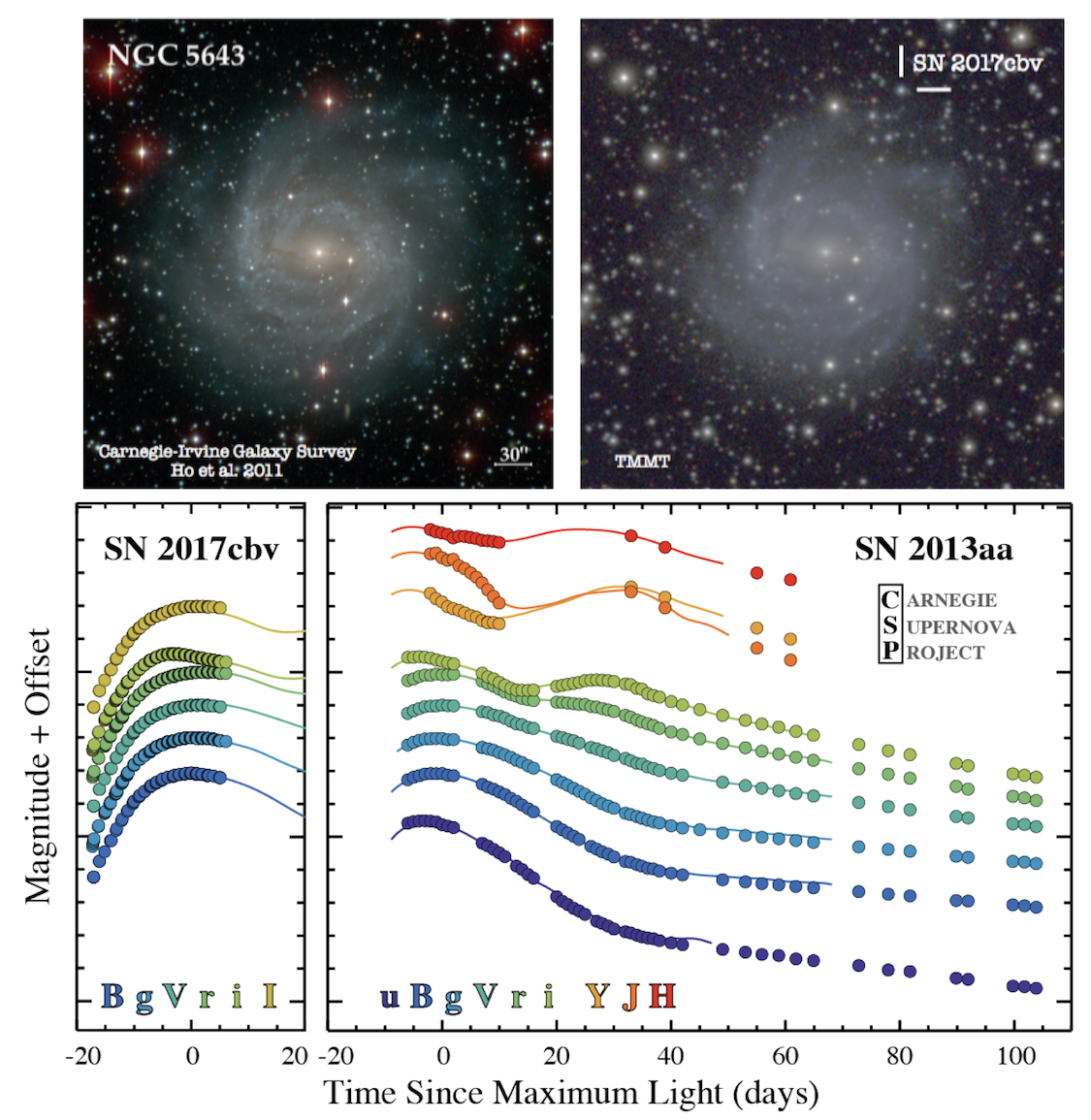}

\caption{ \label{fig:snphot}
Imaging data required to characterize \sn~ for use in a Hubble diagram. 
(top panels) Pre-imaging of NGC\,5643 from the Carnegie-Irvine Galaxy Survey \citep{ho_2011}
as compared to a photometric monitoring image of NGC\,5643 with SN\,2017cbv indicated \citep[imaging using the TMMT;][A. Monson, priv.~communication]{monson_2017}. 
(bottom panels) Daily photometric monitoring of SN\,2017cbv (left) in six photometric bands from a pair of robotic telescopes at Las Campanas Observatory through maximum light. 
This exquisite pre-maximum photometry is enabled by dual-hemisphere monitoring programs and can be compared light curves from the previous SN Ia in NGC\,5643, SN\,2013aa (right). 
The nine-band photometric monitoring from the CSP permits detailed study of the the SN Ia, including direct fits for the extinction law local to the SN Ia.
SN\,2017cbv will be followed for a similar length of time as SN\,2013aa from the ground and space. Small ticks correspond to a luminosity change by 1 mag. Light curves courtesy of A. Monson, M. Seibert, and C. Burns. 
}
\end{figure*}

\subsection{SN~2017cbv: A Recent `local' SN Ia}

On March 10, 2017, a transient source, DLT17u, was discovered in the nearby (D=16 Mpc), face-on barred spiral galaxy NGC\,5643 within the context of the D\textless40 Mpc (DLT40) one day cadence supernova search \citep{2017cbv_atel}. 
The top panels of Figure \ref{fig:snphot} compare a pre- and post-discovery image of NGC\,5643. 

Within 20 hours, a spectrum was obtained for the source that classified it as a SN Ia from the characteristic broad emission lines \citep{cbv_classification}. 
Comparison of the spectrum to a series of templates indicated this object was found at least 2 weeks before maximum light and detailed comparisons to template spectra within the Superfit code \citep{howell_2005} confirmed this was young SN~Ia. 
Based on this classification, the transient DLT17u was renamed to SN\,2017cbv. 

Additional careful photometric monitoring of the object began nearly immediately after its discovery \citep[e.g.,][among others]{cbv_swopeobs} and, due to its brightness (it reached $\sim$ 11.5 mag in $V$), both professional and amateur astronomers have participated in data collection.
It is noteworthy that NGC\,5643 was home to a SN~Ia in 2013, SN\,2013aa \citep{2013aa_atel}, which was extensively monitored by the Carnegie Supernova Project  \citep[CSP][]{freedman_2009,hamuy_2006}.
In contrast to SN\,2017cbv, SN\,2013aa was only classified `a few days before maximum \citep{2013aa_atel}, which limited follow-up observations by other teams. 

\begin{figure*}
\centering
\includegraphics[width=1.0\textwidth]{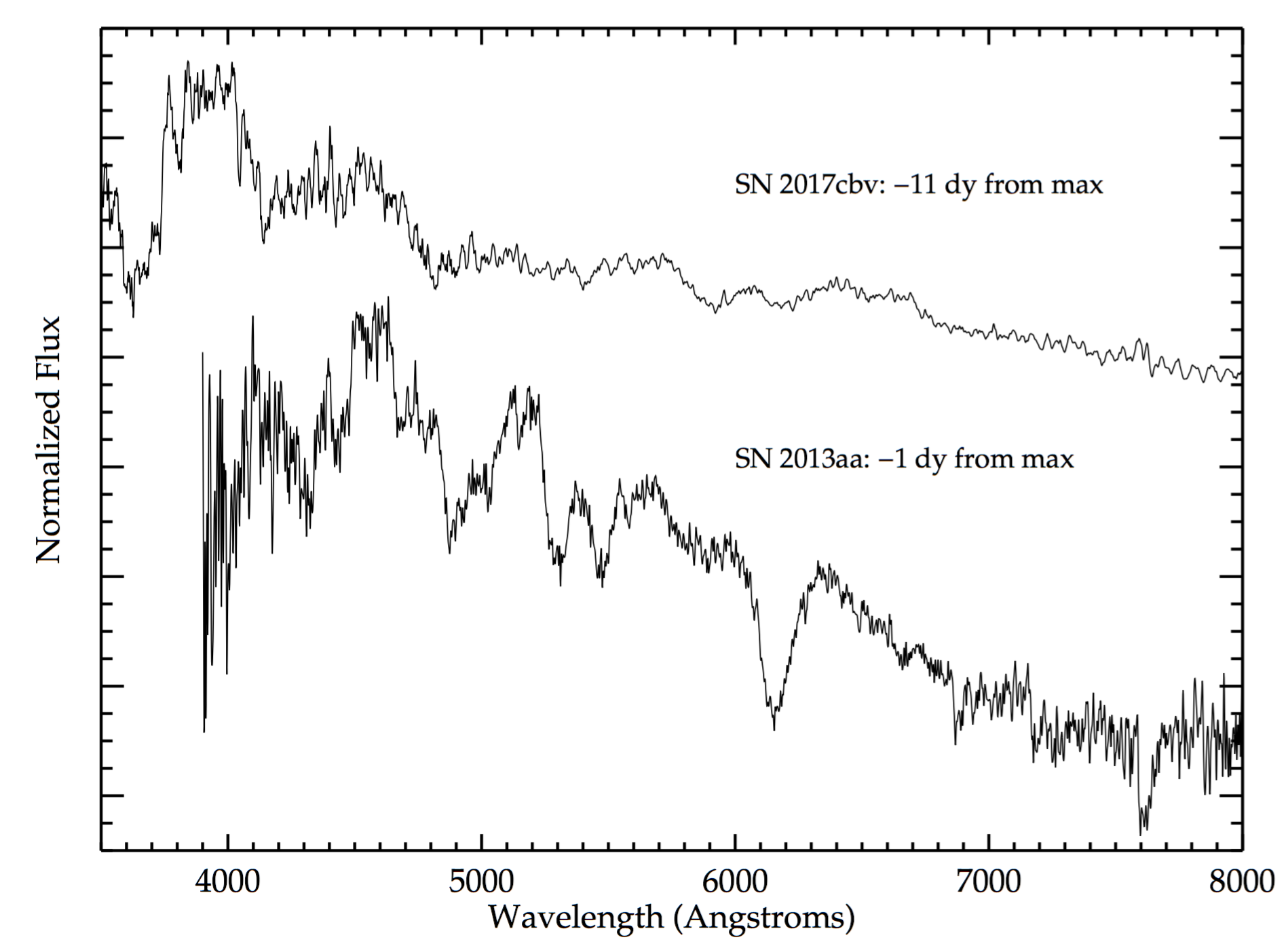}
\caption{ \label{fig:snspec}
Classification spectra for the \sn~ in NGC\,5643.
SN\,2017cbv was classified at least -10 days from maximum and the same day of discovery (top). 
SN\,2013aa was classified -1 day from maximum and three days after its discovery (bottom).
These two spectra are demonstrative of the differences in \sn~ spectra on $\sim$10 day timescales. 
The spectrum for SN\,2017cbv was retrieved from the Transient Name Server (\url{https://wis-tns.weizmann.ac.il/object/2017cbv}) and obtained by \citet{cbv_classification}.
The spectrum for SN\,2013aa was retrieved via the Latest Supernovae \citep{galyam_2013} compilation and obtained by Terry Bohlsen (\url{http://users.northnet.com.au/~bohlsen/Nova/sn_2013aa.htm}).
}
\end{figure*}

In Figure \ref{fig:snspec}, the -11 day spectrum of SN\,2017cbv is compared to the -1 day spectrum of SN\,2013aa. There are marked differences between the spectral evolution over just this 10 day time span, but the broad, often asymmetric, spectral features are characteristic of \sne. 
The ability to find these events early and begin the follow-up process not only aids in the use of a \sn~ for distance determination, but also in revealing their physical evolution. 
A detailed guide to optical spectral classification for all SNe is given in \citet{filippenko_1997} and for \sne~ by \citet{parrent_2014};  in particular, \citet{parrent_2014} demonstrates the spectral evolution with time, for which the spectrum of SN\,2017cbv represents one of the earliest ever observed. 


The photometric monitoring data for SN\,2017cbv and SN\,2013aa are compared in the bottom panels of Figure \ref{fig:snphot}.
Multi-band simultaneous light-curve fits from SNooPy \citep{burns_2011,burns_2014} are also shown for both \sne~ underneath the individual observations.
The key difference is the (i) dense sampling permitted with robotic facilities and (ii) the \textgreater~10 day difference in the starting point of the light curve. 
Utilizing a \sn~ for a distance measurement requires careful fitting of the time of maximum light, which is less ambiguously determined for SN\,2017cbv than for SN\,2013aa.
Moreover, to simultaneously and self-consistently solve for both the total extinction (the combined foreground Galactic extinction and the internal host-galaxy extinction) and the reddening law (e.g., $R_V$) local to the \sn, simultaneous multi-band fitting is necessary. 
Preliminary fitting with SNooPy for both \sne~ suggest that the total extinction for both objects is consistent with the foreground component; this is consistent with the location of either SN~Ia in the outer disk of NGC\,5643, 

The light curves for SN\,2013aa which are typical of the CSP program, show the differences between the light curves observed in blue and red wavelengths. 
At longer wavelengths (longward of $i'$, there is a secondary maximum that occurs around 35 days. 
The strength and timing of
 the secondary maximum is correlated with the decline rate (with the fastest decliners showing no bump), which is suggestive that it has some physical connection with the primary maximum \citep[see discussion in][and their Fig.~1]{phillips_2012}. 
The near-infrared properties of \sne~ have revealed additional peculiar classes not apparently in the optical light-curves. 
For this reason, \citet{phillips_2012} argues that the $Y$ is well positioned to meet the complementary needs of signal-to-noise, insensitivity to dust, well-behaved decline-rate corrections, and simplicity of technology to provide the greatest leverage on the future cosmological measurements using \sne. 

The light curves for SN\,2013aa also demonstrate the length of time for typical SN~Ia follow-up.
More specifically, to determine the stretch-factor for the Phillips-relation, $\Delta m_{15}$, observations must be sufficient to (i) determine maximum light and (ii) to follow the light curve beyond 15 days. 
Both \sne~ display `ideal' stretch factors of $\Delta m_{15} \sim 1$ and, thus, both \sne~ are `normal' and ideal for use as standard candles.
As is demonstrated for SN\,2013aa, the `local' sample is typically followed with a dense cadence for $\sim$40 days (to detect the secondary maximum) and then a less dense cadence there after for as long as the \sn~ can be observed from the ground.
When the \sn~ is no longer detectable from the ground, space-based photometry can be used to put further constraints on progenitor models; SN\,2011fe in M\,101 has been followed 1840 days post its $B$ maximum (though on months long timescales) and observations are planned for at least another 600 days \citep{shappee_2017}. 

The discussion of SN\,2017cbv and SN\,2013aa demonstrates the progress in the level of \sn~ observations in only a few years. 
Dual-hemisphere transient monitoring enabled by robotic, and largely autonomous, facilities permits early discovery of transient objects.
A large-scale networked international community, enabled with state-of-the-art data processing tools, can classify a transient event within a day of its discovery.
Classification, in turn, triggers numerous communities poised for appropriate follow-up, which for the use of SN~Ia for distances requires photometric monitoring for up to 40 days post-maximum. 

\subsection{The `local' Sample in Context}

\begin{figure*}
\includegraphics[width=0.5\textwidth]{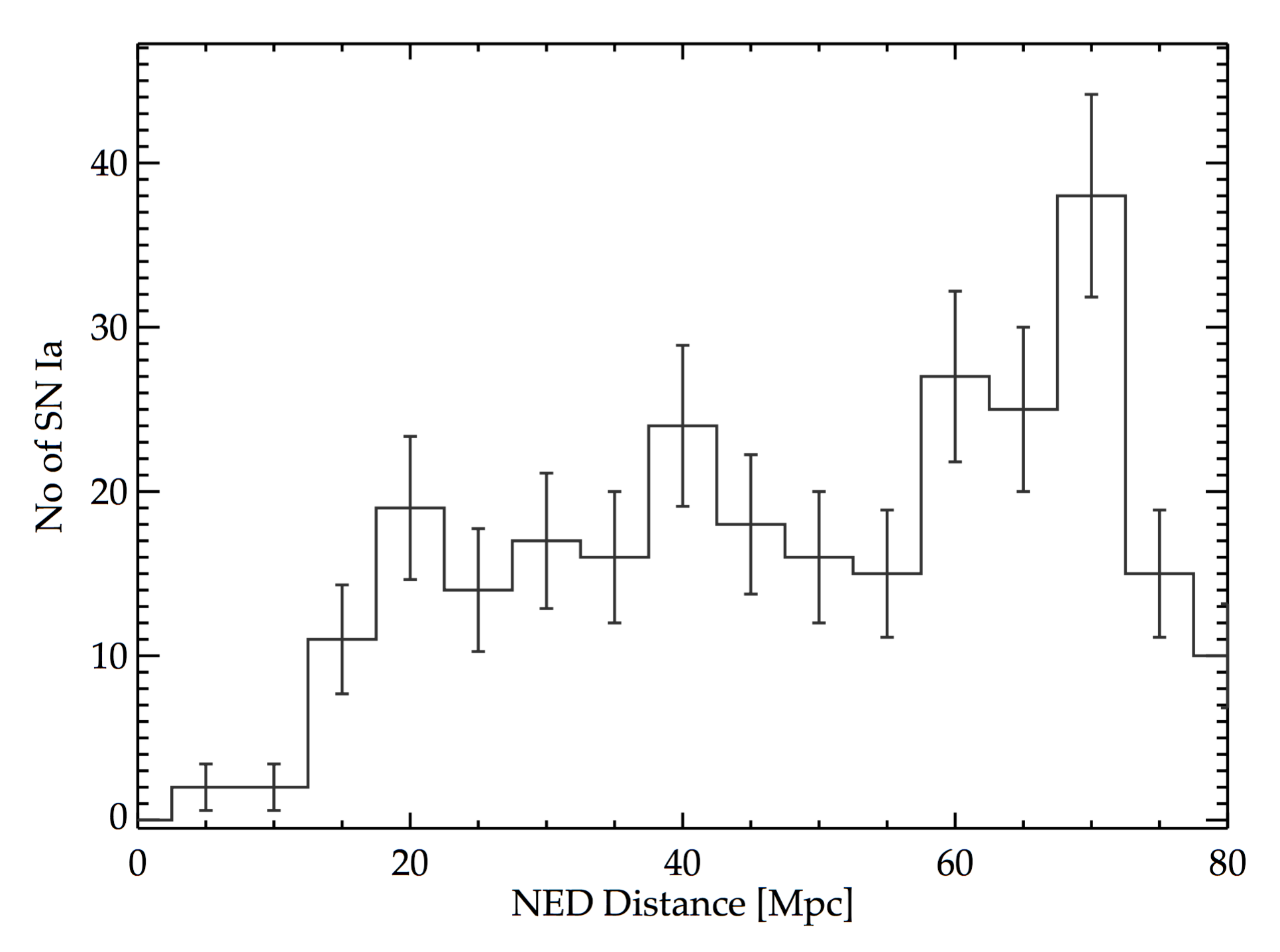}
\includegraphics[width=0.5\textwidth]{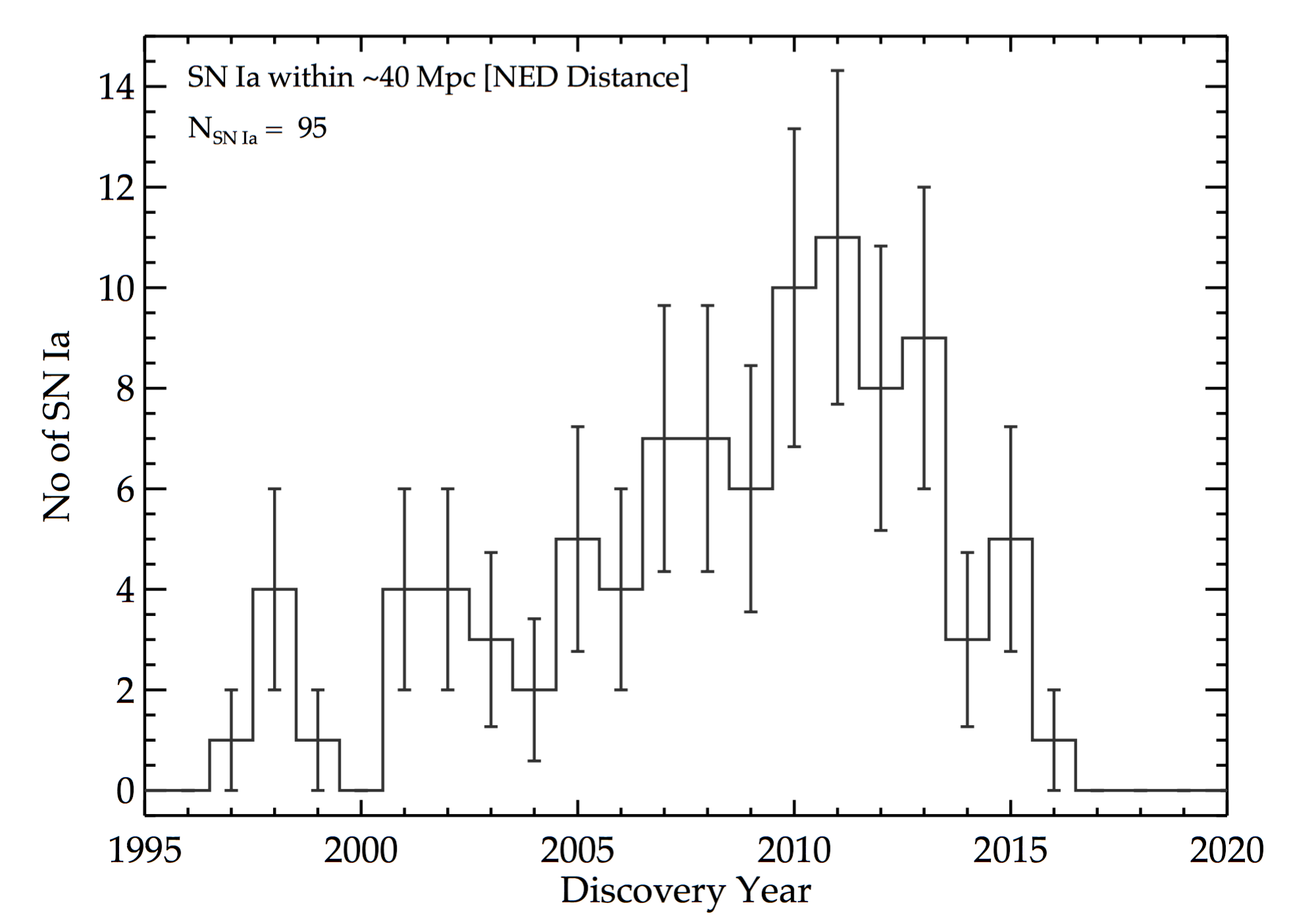}
\caption{ \label{fig:sn_database}
'Local' \sne~ demographics as of as of March 01, 2016.
{\bf (a)}
Number of normal \sne~  discovered since 1995 as a function of the NED Homogenized distance for their host. 
Error bars show the Poisson uncertainty for each bin.
While there are few \sne~ within 20 Mpc, the distribution beyond 20 Mpc is relatively flat.
{\bf (b)}
Number of normal \sne~ discovered per year since 1995 
 within 40 Mpc. 
 There are a total of 95 \sn~ using these distances.  
This suggests that if primary distance techniques can push into this volume,
 a vastly larger number of \sne~ can be independently calibrated to study
the known biases in the \sne~ population.
}
\end{figure*}

The current limitation of \sne~ as {\it absolute} distance indicators is the calibration of the absolute luminosity of \sn. We do not have theoretical predictions of their peak luminosity because there is still continuing uncertainty about their exact nature. Thus they have to be calibrated using other distance indicators, and this is one of the primary purposes for the modern distance ladder. \sne~ are calibrated by Cepheid stars, using the Leavitt Law, which in turn have to be calibrated to an absolute scale. Currently, the Leavitt law is determined by using both Cepheids in the Milky Way and in the Small and Large Magellanic Clouds. Milky Way Cepheids have distances measured geometrically from their parallaxies, either by Hubble Space Telescope or Hipparcos \citep{vanleeuwen2007}, and new \emph{Gaia} masurements, for a much larger number of stars \citep{gaia2017} are now being incorporated into the cosmic ladder. The Leavitt Law in the Large Magellanic Cloud is anchored by eclipsing binaries \citep[e.g.][]{pietrzynski2013}. The statistical error in the distance measurement of a single Cepheid was estimated to be around 0.3 mag \citep{riess_2016}.

However, as of  \citet{riess_2016}, there are in total of 19 SNe Ia host galaxies that can be calibrated via Cepheids and the Leavitt Law. 
Prior to 2016, there were only eight SN~Ia calibrators, which were (largely) the same SN~Ia used in the \emph{HST} Key Project \citep{freedman_2001}.
This implies that roughly $\sim$1 SN~Ia suitable for the calibration sample is discovered per annum, despite SNe~Ia comprising 66\% of the sample of all bright SNe \citep[see][and references therein]{asassn_15}. 
It is of interest to investigate what is limiting the build-up of the SN~Ia calibration sample.

In Figure \ref{fig:sn_database}a, the \sne~ known from the `Recent Supernova' database \citep{galyam_2013}\footnote{Available \url{http://www.rochesterastronomy.org/supernova.html}} as of March 2016 are binned by the homogenized distance to their host galaxy from NASA Extragalactic Database \citep[NED;][]{steer_2017}.
This sample has been cleaned of \sn~ sub-types, but not by extinction (host galaxy or Galactic foreground) or decline rate.
The average number of \sne~ within a single 5 Mpc bin is $\sim$20 \sne, which suggests there are 5$\times$ more \sne~ within the volume accessible to $HST$ Cepheid measurements than are currently in the calibration sample 
(using $\sim$35-40 Mpc as the limiting volume). 

With the dramatic increase in efficiency for transient follow-up and classification described above, is it not unreasonable to attribute this large number entirely to recent discoveries. 
In Figure \ref{fig:sn_database}b the 95 \sne~ within 40 Mpc are binned by their discovery year from 1995 to 2016. 
For reference, the ASAS-SN all-sky project has been in operation since 2013 \citep{asassn_1314}, but only reached peak efficiency in the latter half of 2015 \citep{asassn_15}, shortly before this data were compiled.
From inspection of Figure \ref{fig:sn_database}b, the number of \sne~ discovered per annum fluctuates widely, but, in general, the numbers do not trend strongly with the onset of all-sky monitoring programs (albeit these programs permit unprecedented early detection; e.g., Figure \ref{fig:snphot}). 
Analyses of yearly rates over the next few years, however, will be more informative with programs like ASAS-SN having become dramatically more efficient \citep[e.g.,][]{asassn_16}. 

Of the 95 SN~Ia within 40 Mpc, nearly half ($\sim$40) have sufficient photometric data sufficient to be included in the calibration sample (e.g., similar in spirit to Figure \ref{fig:snphot}; C.~Burns and B.~Shappee, priv.~communication). 
Despite having sufficient data to determine a \sn-based distance, these host galaxies lack robust independently determined distances from the Cepheid-based distance ladder.
The reason for the lack of independent distances is the unsuitability of the host galaxy for Cepheid based distances, either due to its star formation history, an edge-on inclination, or its morphological type.
Thus, utilization of a distance ladder constructed using a standard candle that can be applied to edge-on and non-star forming galaxies has the potential to dramatically improve the size of the calibration sample.
Such a distance ladder is explored using old stellar populations in \citet{beaton_2016}, and these distance methods are described elsewhere in this volume. 
As discussed in the following section, expanding the calibration sample to better reflect the demographics of \sne~ in the larger, Hubble flow sample is an important step for further refinement of \sne~ derived cosmological parameters.

In principle, the use of multiple calibrators for a single galaxy should help to control the systematic errors. For example, the use of galaxies where both SN I and II have been detected would be valuable, but unfortunately, there are only a few of such galaxies, and they lack distance determination based on the Tip of the Red Giant Branch (TRGB) and/or Cepheids. Probably, with the new facilities the number of galaxies with both SN types and a good distance determination will increase.

Apart from Cepheids, NGC 4258, an active galaxy with spatially resolved water maser emission is used as additional distance indicator since this galaxy is also a host for numerous Cepheids. A combination of four geometrical distance calibrators \citep{riess_2016} allowed {\bf a determination of $H_0$ with 2.4\% uncertainty, and this is the uncertainty currently underlying all cosmological measurements done with the use of \sne~ as a systematic error.}

\subsection{In the Hubble Flow}

Our discussion of the `local' \sne~ in NGC\,5643 is characteristic of the attention received by these rare `local' calibrators.
The dramatically larger sample of \sne~ in the Hubble Flow receive less attention in their data collection, but represent a much more complete sample from which to draw inferences on \sn~ as a population. 
More specifically, these are usually only monitored for $\sim$20 days to obtain decline rates and may not have extensive multi-wavelength data or time-resolved spectroscopy. 

There are three main sources of systematic uncertainty that hinder the use of \sne~ and our understanding of the physical phenomena inferred from them \citep[adapted from][]{hicken_2009}:
\begin{enumerate}
\item \emph{photometric accuracy}: large samples of \sn~ light curves are a largely heterogeneous ensemble of photometry coming from multiple programs operating with different sky-footprints and detectors. 
\item \emph{host-galaxy reddening}: \sn~ are discovered in galaxies with a range of host morphologies and are found across these galaxies and thus span a large range of local conditions. Many \sn~ show anomalous $R_{\lambda}$ {\bf extinction parameter, where $R_{\lambda} = A_{\lambda}/E_{B-V}$ is defined by the ratio of the total extinction to the color excess}, but this could be attributed to many factors. Constraining the host-galaxy
 reddening is of paramount importance. 
\item \emph{\sn~ population differences and/or evolution}: the intrinsic variation of the \sn~ population as a function of the stellar populations of the host galaxy, etc. 
\end{enumerate}
The large scale programs described earlier are all aimed at addressing item (1). Item (2) will be discussed in detail later in this volume (see Section~\ref{sec:extIa}). The relevant results for item (3) are summarized in the points to follow.

\noindent \textit{Multi-\sn~ Systems:}~~ There are only two current `local' galaxies that are host to multiple \sn~ in the modern observation era. 
These are NGC\,1316, a massive elliptical galaxy in the Fornax cluster, and NGC\,5643, a spiral galaxy (discussed above). \citet{stritzinger_2010} used the four events in NGC\,1316 and were able to determine that the three normal \sne~ provided consistent distances at the 5\% level and demonstrated that the fourth \sn, a fast-decliner, was discrepant at the 25-30\% level.
The two \sne~ in NGC\,5643 (Figures \ref{fig:snphot} and \ref{fig:snspec}) will be the first time a test of this nature can be performed on a late-type host. 
Moreover, both of these galaxies are sufficiently local that distances can be derived independent of \sne~ to extend such tests in absolute terms (in lieu of differential analyses). 

\noindent \textit{Intrinsic Scatter:}~~Using early data from the CSP, \citet{folatelli_2010} studied the scatter of the best studied \sne~ in their sample; this set included 23 \sne~ with multi-wavelength data similar to that presented for SN\,2013aa in Figure \ref{fig:snphot}. 
A scatter of 0.11 mag (5\%) in distance was reported. 
\citet{folatelli_2010} noted that the dispersion increased with decreasing distance, which implies that much of the observed dispersion could be accounted for with peculiar velocities. 
They discussed additional reasons for the observed trends, concluding that many observational and (potential) intrinsic effects could contribute to the scatter. 
A deeper analysis using a larger set of homogeneous data would be beneficial to understand the intrinsic scatter in \sne~ magnitudes. 

\noindent \textit{Host Mass Bias:}~~\citet{sullivan_2010} find a strong dependence on the mass of the host. 
Massive galaxies are systematically brighter by 0.06 to 0.09 mag (3\% to 5\% in distance; \textgreater 3$\sigma$ difference). 
Such effects will average out for cosmology if the distribution of host galaxy masses remains relatively constant in time, but it is unlikely that this is the case for distant \sne~ samples.

\noindent \textit{Local Star Formation Bias:}~~\citet{rigault_2013} found a dependence of derived \sne~ magnitudes as a function of the local (1 kpc) environment around the \sn~ as probed by H$\alpha$ surface brightness. 
\sn~ in passive (i.e., non-star forming) environments were found to be brighter by 0.094 $\pm$ 0.031 mag (5\% in distance), which is termed the local star formation bias. 
\citet{rigault_2015} revisited this issue using both an independent sample of \sn~ \citep[that of][]{hicken_2009} and using $FUV$ maps to trace the local region. 
\citet{rigault_2015} find that \sn~ in locally star-forming environments are 0.094 $\pm$ 0.025 mag (5\% in distance) fainter; a 3.8$\sigma$ difference.
If the  \citet{hicken_2009} sample is isolated to just the most massive hosts, nearly 50\% of the \sn~ arise from locally passive environments.
The local \sn~ calibration sample is almost entirely star-forming \citep{rigault_2015,riess_2016}.

These effects are complicated and may bias attempts to better understand which galaxies produce \sn. More complex modeling as in \citet{holoien_2017} could be of use to disentangle these effects in the general \sn~ population as well as the continued large-scale study of detailed \sn~ host demographics \citep[e.g.,][and references therein]{asassn_16}. 

Additional systematic error in \sne~ measurements may come from weak lensing effect, particularly for the high redshift sources, above $z = 1.5$. Weak lensing causes additional dispersion in the measured luminosity distances, and the distances are on average effectively larger. \citet{hilbert2011} estimated that this effect introduces systematic error of about 2\% at redshift 1.5, but the appropriate simulations may allow to reduce this effect. This applies also to other measurement methods based on standard candle approach.

\subsection{Progenitors}
\label{sec:SN_progenitors}

Despite their use as cosmological standard candles and their relevance also for galactic chemical evolution, 
consensus about the nature of SN~Ia progenitor systems is still lacking, and this may affect their use as the tracers of the Universe expansion. 

Any model for  SN~Ia progenitors must satisfy a number of observational constraints. The first obvious ones are the 
lack of hydrogen ($\Delta M_H< 0.01 M_{\odot}$) and helium in the spectra, the   
range of values of the peak luminosity and their correlation with the light curve width, and a total kinetic energy of the ejecta 
of $\sim 10^{51}$~erg. 
Our understanding of SN~Ia lightcurves dictates that the peak luminosity is determined by the mass of radioactive $^{56}Ni$, 
whilst the lightcurve width depend on the opacity, determined by iron-group elements 
like $^{56}Ni$, $^{58}Ni$, $^{54}Fe$. These latter two elements are non-radioactive and do not contribute to the peak luminosity.
The temporal evolution of the spectra also allows to reconstruct the radial composition of the ejecta. 
All these observations taken together imply the presence of $\sim$0.1-1.0 $M_{\odot}$ of $^{56}Ni$ in the ejecta (but very little 
in the outer layers), $\sim$0.2-0.4 $M_{\odot}$ of Si, S, Ca, Ar, and less than 0.1 $M_{\odot}$ of n-rich isotopes 
($^{54}Fe$, $^{58}Ni$).

It was recognized very early on (Hoyle and Fowler 1960) that SN~Ia must be thermonuclear explosions in electron degenerate matter, 
pointing to exploding white dwarfs (WDs) as the cause of SN~Ia events. Given that single WDs form with masses below the Chandrasekhar 
mass and are intrinsically stable objects, their evolution must be perturbed by a companion in order to cause the required explosion.
Irrespective of the companion and the trigger mechanism, one can firstly investigate which WD mass+explosion mechanism can be invoked 
to explain  \lq{normal}\rq\,SN~Ia observations \citep[see, e.g., the review by][]{ropke}. 
In terms of WD masses, one can divide the potential progenitors into two broad classes, namely Chandrasekhar and 
sub-Chandrasekhar mass progenitors.

In the Chandrasekhar mass scenario a C-O WD in a binary system approaches the Chandrasekhar mass due to accretion from 
the companion, and C-burning reactions set in at the centre when the density is of the order of $10^9 {\rm g~cm^3}$. 
After a non-explosive initial  \lq{simmering}\rq ~ phase \citep[see, i.e.,][]{lesaffre} 
lasting 1000-10000 yr, that burns only a small amount of C, a thermonuclear runaway starts. The number of 
ignition sparks of the burning front and their distribution are unknown and have a significant impact on the characteristics of the 
explosion. The best agreement with observations is achieved considering a combustion front 
propagating at first subsonically (deflagration) then supersonically (detonation -- Blinnikov and Khokhlov 1986). This so-called 
 \lq{delayed detonation}\rq  ~allows the production of a layer of intermediate-mass 
elements that encompasses almost all of the outer ejecta, as required by observations. The transition {\bf from deflagration to detonation}  
is typically parametrized and not derived from first principles. Variations of the flame ignition configuration and transition 
to detonation can reproduce at least qualitatively the observed luminosity-decline rate relation (Kasen et al. 2009).

In the sub-Chandrasekhar mass scenario \citep[see, i.e.,][]{ww} the helium layer that surrounds the CO core 
increases its mass due to accretion from the binary companion. When the mass of the He-layer reaches a critical value,  
a detonation is triggered by compression. This detonation burns the He-layer and drives a shock wave into the core, triggering 
a detonation of carbon and the explosion of the WD.
In this scenario, a variation of the exploding WD mass could potentially explain the luminosity-decline rate relation.
There are however problems related to the explosive nucleosynthesis, for  
not enough stable Fe and Ni isotopes to match late time spectra are produced 
(because of too low densities at explosion), and too 
much $^{56}Ni$ is made in the He-shell ashes \citep{ropke}. If the 
mass of the He-layer that detonates is $\sim$0.01$M_{\odot}$, as possible for WDs around 1$M_{\odot}$ \citep{bsw}, negligible 
$^{56}Ni$ seems to be produced, although other calculations do not confirm this result \citep{ropke}.

Both these WD mass+explosion mechanisms put forward to explain SN~Ia events need the presence of a companion in a binary system. 
We can traditionally divide the progenitor systems into two broad classes, namely 
single-degenerate (SD) and double-degenerate (DD) systems.

In a typical SD system \citep{wi} the WD accretes H from a non electron-degenerate 
companion. Hydrogen is burned to He first, then to C, and the WD mass increases to the Chandrasekhar limit.
As shown by \citet{nk}, stable H-burning is possible only for a finely tuned accretion rate of around $ 3 \times 10^{-7} M_{\odot} 
{\rm yr^{-1}}$. Lower rates ($<10^{-8} M_{\odot} 
{\rm yr^{-1}}$) produce a Nova event, which expels the ashes of the accreted material and possibly even some of the original WD material \citep[e.g.][]{yaron05}. 
This may not be the case at intermediate rates, where recurrent Novae systems retain some of the accreted mass \citep[e.g.][]{hk}. 
 If the accretion rate is too large, the WD is expected to expand to giant-like dimensions, and engulf the donor star in a common envelope which inhibits further accretion onto the WD.  
\citet{hachisu96} have proposed that an optically thick wind emerges from the WD 
that stabilizes the mass transfer. In this way the mass accretion may continue, but at a reduced rate. 
Generally, there is only a narrow range of accretion rates in which the WD can retain the matter and grow in mass.

If the non-degenerate companion is a He-star (produced by a previous common envelope episode) the WD can accrete He  
leading to either a Chandrasekhar or a sub-Chandrasekhar explosion \citep[see, i.e.,][]{wk, pty}, 
depending on the accretion rate. Accretion rates of the order of $\approx 10^{-8} - 10^{-9} M_{\odot} 
{\rm yr^{-1}}$ lead to a He-detonation and a sub-Chandrasekhar explosion; in case of rates of the order of 
$\approx 10^{-6} M_{\odot} {\rm yr^{-1}}$ steady He-burning on the WD surface can produce a Chandrasekhar mass WD that 
ignites C in the core.

In a DD system \citep{ty} two WDs orbit each other and the system loses energy through gravitational waves emission.
As the two components come closer and closer, eventually the least massive object fills its Roche-lobe and mass transfer sets in, 
with the more massive WD accreting C-O from an accretion disk/torus. If this WD can reach the Chandrasekhar mass a delayed-detonation 
will ensue. It is however uncertain whether this happens; even in case of a system where the combined mass 
of the two WDs exceeds the Chandrasekhar limit, an off-centre carbon deflagration followed by 
conversion to an O-Ne-Mg WD seems to be a likely outcome. Nearing the Chandrasekhar mass the  
O-Ne-Mg WD undergoes an electron induced collapse that leads to the formation of a neutron star.
An alternative path that could lead to a SN~Ia explosion is a so-called violent merger  \citep[see, i.e.,][]{pakm10}. 
At the merger, an accretion stream can dynamically produce a high temperature when hitting the WD surface. If high enough, 
this leads to the WD explosion according to the sub-Chandrasekhar double-detonation scenario.
This scenario seems however efficient only for systems whose primary WDs are massive ($> 0.8-0.9 M_{\odot}$) and the mass ratio is high $>0.8$. 
\citet{pakmor13} find that the presence of helium facilitates the ignition.  

Despite the enormous efforts of the astronomical community, there has not yet been a convincing identification of a SN~Ia progenitor \citep[][]{li11, nielsen12}. 
Furthermore, none of the proposed formation channels are clear matches to the observational constraints. Due to the uniformity and continuity of the observed SN~Ia 
properties, there has been a long-time focus  to find a single formation channel to explain all properties. However, from the diversity and correlations among the 
spectral properties of SNe~Ia and their host galaxies, it is also possible that some combination of formation channels is operating.

An advantage to the SD channel is that it naturally explains the uniformity in luminosities as all WDs explode at the Chandrasekhar mass in this channel. On the other 
hand, the masses of the merged remnants in the DD channel display a broader range ($\sim1.4-2.0 M_{\odot}$). It is usually assumed that the DD systems with super-Chandrasekhar 
masses  are responsible for the super-luminous SN~Ia. Another expected signature of the SD channel, is that the companion is expected to survive the explosion and with an 
anomalous velocity, rotation, spectrum or composition. 
No such object has been identified unambiguously, for example in Tycho's SN remnant \citep{ruiz04, kerzendorf09, schaefer12}.  
Also, any mixing of the SN ejecta with that of the companion star during the explosion, has not been observed conclusively \citep{leonard07, garcia12}. 
In a minority of cases, variable NaD absorption has been detected in the SN spectra, which has been interpreted as circumstellar material from the companion in the SD 
channel \citep{patat07, sternberg11}. 
This signature, however, can also arise in the DD channel from a post-merger, pre-explosion wind \citep{shen12}. 
Furthermore, one expects soft X-ray emission from SD systems produced by the steady H-burning on the WD surface, but  
not enough emission has been observed from resolved nor unresolved sources \citep{gb, distefano10}.
The emission might be shielded within optically thick outflows and reprocessed to UV-emission \citep{hachisu10, nielsen13, wheeler13}, but so far a population of these 
sources has not yet been detected \citep{woods13, lepo13}.
Lastly, SN2011fe is a recent and close SN~Ia that is well observed in optical, radio and X-ray wavelengths. These observations have ruled out most types of donor stars 
in the SD channel, and therefore favor a DD progenitor \citep[e.g.][]{chomiuk13}; see also \citet{maoz14} for a review on the observational clues to the SN~Ia progenitor 
problem. 

Additional constraints on the SN~Ia progenitor channels are provided by the observed delay-time-distribution (DTD) of SN~Ia events. The DTD is defined as the time 
interval between a star formation episode and the explosion of the related supernovae.
It illustrates the distribution in evolutionary timescales of the SN~Ia progenitor, which vary for the different progenitor scenarios.
The DTD has been studied observationally in a range of surveys, environments and redshifts, 
and the emerging picture is remarkably coherent \citep[for a review, see][]{maoz12}.
Type Ia SNe occur in young and in old stellar populations, reaching delays as long as a Hubble time. The SN~Ia rate peaks at short delays of $<1$\,Gyr and declines 
at longer delay times. 
The DTD is best described by a power-law with an index of about $-1$ for delays of $1 < t < 10$\,Gyr. 
At shorter delays, there is still some uncertainty regarding the precise shape of the DTD. 
Despite the consistency in the shape of the DTD from a variety of methods, the normalization or time-integrated DTD
shows variations between rates in different galaxy types. SN~Ia rates in galaxy clusters are found to be a factor $\sim$5 higher than 
rates based on volumetric galaxy samples. Further research is needed to test if this is due to differences in the methods and samples, or 
if there is indeed an enhancement of the SN~Ia rate in cluster environments \citep[e.g.][]{maoz12b,maozgraur17}. 

The DTDs for different progenitor channels have been estimated by analytical approaches and by modeling the evolution of binary populations \citep[see][for a review]{wang12}.
For the double degenerate channel, there is a good agreement on the DTDs \citep[e.g.][]{nelemans13}. 
The DD model gives rise to delay times ranging from a few Myr up to a Hubble time. Within these delay times, the DTD shape is a 
continuous power-law with a slope of roughly -1, which is remarkably similar to the
observed DTD. The time-integrated DTD from the DD channel is compatible with the lower limit of the observed rate.
The DTD from double WDs that undergo a violent merger is similar to that of the classical DD channel \citep{ruiter13}. 

For the single degenerate channel, there is a large diversity in the predicted SN~Ia rates spanning over several orders of magnitude \citep[e.g.][]{nelemans13}.
Differences arise due to ill-constrained aspects of binary formation and evolution, most importantly
the accretion efficiency of WDs \citep{bours13} and common-envelope evolution \citep[e.g.][see \citet{ivanova13} for a review]{claeys14}.
The predicted rates of the SD channel tend to be (far) below that of the DD channel and the observed rate. 
Typically, the delay times from the SD channel range from a few-hundred Myr to a few Gyr. The models show that SD DTDs have a sharp drop after a few Gyr,  in contradiction with the observed DTD.
The drop can be understood by the limited range of donor masses ($\sim 2-3 M_{\odot}$) that transfer matter to the WD at the necessary rates to ensue WD mass growth. 

If the donor star is a He-star, the delay times are very short i.e. $\sim$50--200Myr \citep{wang09}, 
due to the fact that He-stars evolve from more massive main-sequence stars with shorter evolutionary timescales than in the standard SD channel.

An adaptation to the SD channel has been proposed that resolved some of the observationally discrepant signatures. In the so-called 'spin-up/spin-down'-model \citep{justham11, distefano11}, WD rotation due to accretion onto the WD is invoked to support against the collapse and ignition of the WD. The time that it takes 
for the WD to spin down and explode leads to a prolonged delay time. This may solve the issue that the classical SD progenitors do not have long delay times that are observed. 
Furthermore, if the timescale for spin down is sufficiently long, any traces of the mass transfer and even the companion could disappear. 
Lastly, the 'spin-up/spin-down'-model also allows for the possibility of super-Chandrasekhar SN~Ia.

\subsection{Extinction/Color Corrections for Type-Ia Supernovae}
\label{sec:extIa}


A common perception exists that the dominant parameter driving the
standardization of SNe~Ia for measuring distances
is related to the decline-rate or light-curve shape \citep{phillips_1993,riess_1995,hamuy_1996,perlmutter_1997}.
However,
corrections based on the color of the SN~Ia are at least as important
\citep{riess_1996,phillips_1999}. A variety of approaches
has been used for this correction, and these differences ultimately
relate to the cause of color variations in SN~Ia, something that has
been difficult to pin down.

All SN~Ia luminosity distance measurements must correct for extinction
by dust in the Milky Way, typically using the dust maps of \citet{schlegel_1998}, 
or more recently, \citet{schlafly_2011}, to
provide an estimate of the reddening $E(B-V)_{\rm MW}$ along the line
of sight. This reddening is converted into an extinction in the
observed passbands using the dust law of \citet{cardelli_1989}, with
updates from \citet{odonnell_1994} and \citet{fitzpatrick_2007},
typically assuming an extinction law parameter $R_V \approx 3.1$ as
found in the diffuse ISM. Care must be taken to account for the
time-evolving spectral energy distribution (SED) of the
supernova. Uncertainties in the Milky Way extinction correction are
correlated across all observations of an individual supernova, and
systematic uncertainties in the Milky Way extinction correction can
have important effects on parameter inferences from a supernova
sample. Nonetheless, hereafter we discuss SN~Ia extinction or color
corrections under the assumption that the Milky Way component has been
properly removed.

In analogy to Milky Way extinction, the light from distant SN~Ia is
extinguished by dust in the supernova host galaxy (which acts on the
rest-frame light in contrast to the Milky Way extinction which acts
after the light has been redshifted to the observer frame). One
approach to this correction is to assume that the intrinsic colors (or
SED) of a SN~Ia are strictly determined by its light curve shape,
i.e. that SN~Ia are intrinsically a one-parameter family. Color
variation among SN~Ia is then ascribed to intrinsic stochasticity
(random scatter around the nominal color) and extrinsic reddening by
host-galaxy dust. This is the approach used by the Multicolor Light
Curve Shape method \citep{riess_1996,riess_1998} and as extended to
MLCS2k2 \citep{jha_2007}, as well as adaptations of the
$\Delta m_{15}$ method \citep{phillips_1999,burns_2011}. Even
if the intrinsic SN~Ia light curves are described with a more complex
parameterization, extrinsic host-galaxy extinction can be treated
separately, as in the BayeSN methodology \citep{mandel_2011}.

Applying this method gives a surprise: Hubble diagram residuals are
minimized when the ratio between the inferred
extinction $A_V$ and color excess $E(B-V)$ has a value $R_V \approx$
1--2, significantly lower than the canonical $R_V \approx 3.1$ 
\citep[e.g.,][]{conley_2007}. This could be interpreted as saying the typical
dust in SN~Ia host galaxies has smaller grains on average than Milky
Way dust, with significantly more reddening for a given amount of
extinction.

Indeed, we find strong evidence for ``weird'' low-$R_V$ dust in
heavily extinguished SN~Ia (e.g., host reddening $E(B-V) \geq 1$
mag). This has been inferred using near-UV through near-infrared
observations of reddened objects like SN~1999cl, 2002cv,
2003cg, 2006X, and SN~2014J \citep{elias_2006,elias_2008,krisciunas_2006,krisciunas_2007,wang_2008,burns_2014,amanullah_2014,amanullah_2015,foley_2014,brown_2015b,gao_2015}. Independent, corroborating evidence comes from linear
polarization measurements; dust scattering imprints a wavelength
dependence to the continuum polarization fraction that reflects the
grain size distribution and $R_V$. For several of these
heavily-reddened objects, the wavelength of the continuum polarization
peak implies $R_V \lesssim 2$ \citep{kawabata_2014,patat_2015}.

An intriguing possibility is that this strange dust is located near
the supernova, in the circumstellar environment of the progenitor
system, perhaps with multiple scattering playing a role \citep{wang_2005,goobar_2008,foley_2014}. This could help explain why such dust
is not evidently seen in the Milky Way ISM, for example. Indeed, some
of these heavily-reddened SN~Ia show time-variable absorption lines
(Na I D, K I) in high-resolution spectra, interpreted as arising from
circumstellar gas interacting with the SN radiation field \citep{patat07,blondin_2009,graham_2015}.

However, the preponderance of the evidence suggests this low-$R_V$
dust is \emph{interstellar}. Many of these SN~Ia show light echoes,
with dust sheets $>$ 10 pc away from the supernova 
\citep[i.e., not circumstellar][]{wang_2008b,crotts_2008,crotts_2015,maeda_2015,yang_2017}. Moreover, the dust seems to
correlate best with diffuse interstellar bands (DIBs) rather than
circumstellar gas absorption (Phillips et al.~2013) and moreover, for
SN~2014J most of the gas absorption originates on interstellar, not
circumstellar scales \citep{ritchey_2015,jack_2015,maeda_2016}.

Because of the difficulties in arriving at a physical understanding of
the relationship between the observed luminosity and color of SN~Ia, a
leading approach is to make this an entirely empirical correction,
analogous to the light curve shape correction. \citet{tripp_1999}
suggested a simple two-parameter empirical correction, with the SN~Ia
$B$-band luminosity linearly regressed against a light curve shape
parameter and a color parameter. The SALT/SALT2 \citep{guy_2005,guy_2007} and SiFTO \citep{conley_2008} light curve fitters adopt this
model. The single color coefficient (called $\beta$ in these models)
is again found to be significantly lower ($\beta \approx 2.5$) than
what would be expected by standard dust 
($R_V = 3.1 \leftrightarrow \beta = 4.1$).

These fitters are aimed primarily at cosmological SN~Ia samples,
i.e. objects typically with low extinction, not the heavily reddened
SN~Ia for which there is more direct evidence of unusual
dust. Thus, it is not clear whether the low values of $\beta$ in
cosmological samples arise from the same kind of unusual dust (just
less of it). For example, measurement uncertainties would lead to
color ``noise'' that was uncorrelated to SN luminosity; if this noise
level were comparable to the dust reddening (typically $E(B-V) \leq
0.3$ mag for these ``cosmological'' objects), the effective (observed)
value of $\beta$ would be lowered \citep{dellaValle_1992}. Similarly, if the SN~Ia had intrinsic color variations, perhaps
also correlated to luminosity with some $\beta_{\rm intrinsic}$, the
effective $\beta$ would be intermediate compared to $\beta_{\rm dust}$ \citep{mandel_2016}.

There is positive evidence that intrinsic color variations and the
error model for SN~Ia colors may be playing a role. By using
spectroscopic indicators that are independent of color (line
equivalent widths or velocities), it is possible to ascribe an
intrinsic color to an individual SN~Ia, and thus determine the color
excess (observed minus intrinsic). Correlating the SN~Ia luminosity
against this color excess (rather than the observed color) yields an
extinction law consistent with normal dust \citep[$R_V \simeq 3$; $\beta
\simeq 4$][]{chotard_2011,foleyKasen_2011,foley_2011,sasdelli_2016} for low-extinction objects. Similarly,
an intrinsic variation ``color-smearing'' model, combined with a
normal dust law, can explain the luminosity-color relation and its
scatter better than a single linear model can \citep{scolnic_2014,mandel_2016}. Nonetheless, strange (strongly-reddening,
low-$R_V$) dust is still necessary for highly-reddened SNe~Ia \citep{mandel_2011}.

The emerging picture to explain the relationship between SN~Ia
luminosity and color has grown quite complex, including intrinsic
color variations that may depend on light-curve shape, spectral
features, or even host galaxy environment, normal Milky Way-like dust
(in the SN host galaxy and in the Milky Way), and strange, likely
interstellar but possibly also circumstellar, low-$R_V$ host-galaxy
dust. Achieving the most precise and accurate distances from SN~Ia
will require modeling and disentangling all of these effects, and
perhaps more. These effects are likely to be more severe for high-redshift SN Ia science, for which there might be significant evolution relative to the Hubble-flow sample, and less severe in measuring $H_0$ from SNe Ia, where the calibrator sample  is better-matched.  For example, using SNe Ia as near-infrared standard candles, less sensitive to extinction corrections than in the optical, does not significantly changed the derived value of $H_0$ \citep{dhawan2017}.

\subsection{Summary}
Future discoveries with SN~Ia cosmology are now built on a framework of large-scale, homogeneous follow-up that unite teams that (i) find, (ii) classify, and then (iii) characterize the SN~Ia. 
There are differences in the follow-up for the `local' sample of SN~Ia, which are ultimately used to set the SN~Ia absolute luminosity, those SN~Ia discovered in the Hubble Flow (to z$\sim$0.1), which are used to determine the Hubble constant, and those found at higher redshifts, which are used for understanding the acceleration of the Universe.
The low-redshift samples, however, remain the lens for understanding the physics of SN~Ia and projecting that back in time; of key interest being how SN~Ia properties may evolve with the decreasing progenitor metallicity over cosmic time and the apparent star-formation biases.
Thus, further development of the `local' sample and bolstering of the low-redshift sample remains as critical as the original questions posed by \citet{shapley_1919} as to the size of the Universe.

\section{Supernovae II}
\label{sec:SNII}

Type IIP SNe can be used as ``standardized'' candles to estimate distances well within the Hubble flow, with a rms precision of the order of $10-15 \%$. This occurence makes them interesting cosmological distance indicators, since they are produced by different stellar populations than type Ia SNe, making them useful sanity check of the type Ia SNe-based results. More generally, it has demonstrated that all type II (IIP + IIL) SNe can be used as standardized candles \citep[e.g.,][]{deJaeger2015, deJaeger2017}. Here the current status of the type IIP SNe standardized candle method is reviewed, pointing out the need for a calibration based on primary distance indicators.

\subsection{Introduction}
\label{intro}
In the classical classification scheme \citep{Filippenko1997}, type IIP SNe are characterized spectroscopically by strong hydrogen emission features, with strong P-Cygni profiles; and photometrically by a long plateau, lasting on average $\sim 80$ days, followed by a sudden drop in luminosity and a subsequent settlement on the radioactive tail.

From the physical point of view, both theoretical \citep[e.g.,][]{Grassberg1971,Litvinova1983,Utrobin2008,Pumo2011} and empirical investigations \citep[e.g.,][]{Smartt2009paper} show that type IIP SNe are the product of a core collapse of a small to moderate massive progenitors, typically red supergiants (RSG). Interestingly, while on the basis of the empirical models \citep{Heger2003,Walmswell2012} we expect that type IIP SNe are the final fate of progenitors of masses between $8 M_\odot$ and $30 M_\odot$, empirical evidence found progenitors only in the range $8-17 M_\odot$. This discrepancy has been dubbed the ``RSG problem'' \citep{Smartt2009review}. However, some claims in the recent literature suggest higher mass limits \citep[e.g.,][]{DallOra2014}, while independent studies of the massive star populations in the Local Group found RSGs with masses up to $\sim 25 M_\odot$ \citep{Massey2000,Massey2001}.

The characteristic photometric plateau is supported by the hydrogen recombination front, which recedes in mass as the photosphere expands and cools down, producing a constant luminosity. As the density lowers down to values in correspondence of which the atmosphere becomes transparent, a sudden drop ($\sim 30$ days) in luminosity of several magnitudes is observed, and the light curve is subsequently powered only by the radioactive decay of $^{56}$Co to $^{56}$Fe. In this phase, the SN luminosity depends on the amount of $^{56}$Ni synthesized in the explosion \citep[e.g.,][]{Weaver1980}.

\begin{figure}
 \includegraphics[width=0.6\textwidth]{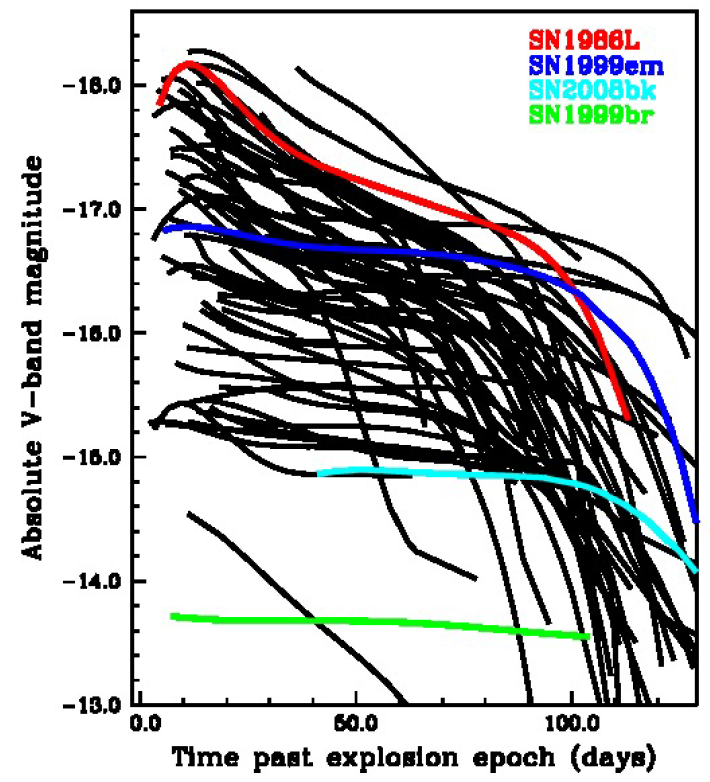}
\caption{Absolute V-band magnitudes of a sample of $60$ type IIP SNe, as published in \citet{Anderson2014}. Reproduced by kind permission of the authors.}
\label{fig:Anderson}       
\end{figure}

Since type IIP SNe are produced by a variety of progenitors, with different masses and chemical compositions, the observed features (luminosity at the plateau, length of the plateau, kinetic energy of the ejecta, amount of synthesized $^{56}$Ni) can be very different from SN to SN \citep[see Fig.\ref{fig:Anderson}, from][here reproduced by kind permission]{Anderson2014}. Indeed, absolute plateau magnitudes range typically from $M_V = -15.5$ mag to $M_V = -18.5$ mag, initial velocity of the ejecta are of the order of $1-2 \times 10^4$ km sec$^{-1}$, and initial photospheric temperatures are of usually around $1-2 \times 10^4$ K.

Nevertheless, interestingly and physically motivated relationships between observable quantities, such as between the linear radius (estimated from the velocity curve) and the angular radius (estimated by fitting a black body to the observed fluxes at different epochs); or between the luminosity at the middle plateau and expansion velocity, allow us to use them as ``standardized'' distance indicators. The first method is known as the expanding photosphere method \citep[EPM,][]{Kirshner1974}, while the second has been dubbed the ``standardized candle method'' \citep[SCM,][]{Hamuy2002}. Quite interestingly, the two methods can give consistent results, up to cosmological distances \citep[e.g.][]{Gall2016}.

\subsection{Physical basis of the EPM and SCM method}
EPM is actually a variant of the Baade-Wesselink method, able to produce very accurate results \citep[see][]{Bose2014}. The method requires the measurement of the temperature of the expanding stellar envelope, and of the envelope radius, which in turn comes from measurement of time since explosion and the expansion speed from the Doppler shift of the lines. Thus, it allows to measure the absolute luminosity, and finally to obtain the luminosity distance in a direct way. Type II SNe are intrinsicaly bright, so this method is able to provide distance estimates up to cosmological distances, independently from the adopted distance ladder, thus providing an independent check of the results obtained, for example, with the type Ia SNe, and it can be applied at any phase. However, it is observationally demanding, since it requires multi-band photometric data and good quality spectra. Moreover, some modeling is needed, especially in correctly estimating the dilution factor of SNe atmospheric models respect to a pure black body. A further improvement of the EPM is the spectral fitting expanding atmosphere method \citep[SEAM,][]{Mitchell2002}, based on full NLTE atmospheric code. It can give quite accurate and precise results (e.g. Baron et al. 2004, Bose and Kumar 2014), but it is computationally intensive and requires high S/N spectra at early phases.

SCM was introduced by Hamuy \& Pinto in 2002 \citep[][hereafter HP02]{Hamuy2002}  as an empirical correlation between the SN luminosity and the expansion velocity of the ejecta. They calibrated their luminosity-velocity relation at day $50$ in the $V$ and $I$ photometric bands, which corresponds to a middle-plateau phase for most of the type IIP SNe. Subsequently, \citet{Kasen2009} provided theoretical basis on the SCM. As a matter of fact, SCM is a simple recasting of the Baade-Wesselink method. Indeed, since the expansion is homologous (and therefore the velocity is proportional to the radius), the luminosity $L$ can be written as $L = 4 \pi v^2_{ph} t^2 \zeta^2 T^4_{ph}$, where $v_{ph}$ and $T_{ph}$ are the photospheric velocity and temperature, respectively; $\zeta$ is a dilution factor, which accounts for the departure from a perfect blackbody, while $t$ is the reference epoch. Now, $t$ can be arbitrarily chosen ($50$ day by construction), and $T_{ph}$ is a good proxy of the temperature $T_H$ of the hydrogen recombination, $ T_h \approx 6000$K, nearly a constant along the plateau. Finally, the dilution factor $\zeta$ can be estimated from NLTE models, but it is a strong function of the luminosity, and it can be absorbed in the exponent. It should be explicitly noted that SCM does not need to be applied necessarily at day $50$, and that similar relations are valid all along the plateau. However, at epochs earlier than day $30$ the ejecta temperature is too high, and probably the approximation $T_{ph} \sim T_h$ is not valid. Moreover, the atmospheric velocity curve is rapidly changing during the first $40-50$ days, and a $10-15$ days uncertainty in the explosion epoch can reflect in a substantial bias in the velocity curve.

\subsection{Current calibrations: an overview}
The first calibration was given by HP02 that, on the basis of $17$ literature type IIP SNe, $8$ of which well embedded in the Hubble flow, derived a relation in the $V$ and the $I$ band as a function of the photospheric velocity and of the redshift. Their calibration showed a scatter of the order of $9\%$, comparable with the precision of $7\%$, typical of type Ia SNe. However, an estimate of the absorption is needed, and this is usually a thorny problem when dealing with SNe. The problem was faced by \citet{Hamuy2003}, where the absorption was estimated on the basis of observed $(V-I)$ colors. Indeed, when it is assumed that the intrinsic end-of-the-plateau $(V-I)$ color is the same for all the type IIP SNe and it is a function of the photospheric temperature only, a possible $(V-I)$ color excess is due to the host galaxy extinction. The underlying physical assumption is that in type IIP SNe the opacity is mainly caused by the e$^-$ scattering, so that they reach the same hydrogen recombination temperature as they evolve. However, some discrepancies are obtained, probably due to metallicity variations from one SN to the other. Moreover, as pointed out by Nugent et al. (2006), a color-based extinction correction is impractical for faint (i.e. basically distant) SNe, since it would require a continuous monitoring to catch the end of the plateau, before the luminosity drop.

A subsequent calibration was then proposed in 2006 by \citet{Nugent2006} (N06), where an extinction correction was determined from the rest-frame $(V-I)$ color at day $50$, adopting a color-stretch relationship, as done in Ia SNe studies. A standard $R_V = 3.1$ dust law was used. Moreover, since at moderate redshifts the weak Fe~II $\lambda 5169$ could be hardly measured (because they can be redshifted into the OH forest), they explored the practicality of stronger lines, such as the H$\beta$. Also, they derived an useful empirical relation to scale the observed Fe~II $\lambda 5169$ velocity at a given epoch, to the reference $+50$ day. They obtained the first Hubble diagram at cosmologically relevant redshifts ($z \sim 0.3$) with a rms scatter in distance of $13 \%$, which is comparable with the rms scatter obtained with Ia SNe.

Poznanski and coworkers \citep[][hereafter P09]{Poznanski2009} used a fitting method similar to those adopted by N06, but taking the extinction law as a free paramenter. This procedure yielded a mild total-to-selective absorption ratio $R_V = 1.5$ and, after discarding a few outliers with faster decline rates, they finally obtained a scatter of $10 \%$ in distance.

Subsequently, \citet{Olivares2010} adopted as a reference epoch a ``custom'' $-30$ day from the half of the luminosity drop, to take into account the different length of the plateau phase from SN to SN. By allowing $R_V$ to vary (and confirming a low $R_V = 1.4 \pm 0.1$), they found again that SCM can deliver Hubble diagrams with rms down to $6\%-9\%$. Interestingly, after calibrating their Hubble diagrams with the Cepheid distances to SN 1999em \citep{Leonard2003} and SN 2004dj \citep{Freedman2001}, they obtained a Hubble constant in the range $62-105$ km s$^{-1}$ Mpc$^{-1}$, but with an average value of $69 \pm 16$ km s$^{-1}$ Mpc$^{-1}$ in the $V$-band, and similar values in the $B$ and $I$-bands. The large scatter reflects the fact that only two calibrating SNe were employed, but the average value is very close to our most precise estimate of the Hubble constant, $H_0 = 73.24 \pm 1.74$ km s$^{-1}$ Mpc$^{-1}$ \citep{riess_2016}.

\citet{DAndrea2010} adopted $K$-corrections to determine rest-frame magnitudes at day $50$ for $15$ SDSS II SNe, spanning a redshift range between $ z = 0.015$ and $z=0.12$. They also took into account the rest-frame epoch $50 (1+z)$. Their best-fit parameters differed significantly from those obtained by P09, and they attributed the discrepancy to the fact that their SNe sample could be intrinsically brighter than those of P09. Moreover, they concluded that a major source of systematic uncertainty in their analysis was probably due to the difficulty of accurately measuring the velocity of the Fe~II $\lambda 5169$ feature, and to the extrapolation of the velocity measured at early epochs to later phases. Finally, they warned that the template database should be extended, in order to perform a reliable $K-$correction.

\citet{Maguire2010} extended the SCM to the near-infrared bands, since at those wavelengths both the extinction and the number of spectral lines are lower. The latter aspect implies that NIR magnitudes are less affected by differences in strength and width of the lines (i.e. less sensitive to metallicity effects), from SN to SN. Even though their adopted sample contained only $12$ SNe, they demonstrated that using $JHK$ magnitudes it is possible to reduce the scatter in the Hubble diagram down to $0.1-0.15$ mag, with the error in the expansion velocity being the major source of uncertainty.

\subsection{Discussion and final remarks}
The extragalactic distance scale up to cosmological distances is intimately connected with type Ia SNe, and through type Ia SNe the acceleration of the Universe was discovered \citep{Perlmutter1999,riess_1998,Schmidt1998}. At the present time, current facilities allow us to detect and study type Ia SNe up to $z \sim 1.9$ \citep{Rubin2013,jones_2013}, and recently up to 2.3 \citep{riess2017}, while the next generation of extremely large telescopes will allow us to study type Ia SNe up to $z \sim 4$ \citep{Hook2013}. At high $z$, however, the number of type Ia SNe may significantly decrease, due to the long lifetimes of their progenitors. Alternatively, the ubiquitous type II (core-collapse) SNe could be an appealing choice to probe further cosmological distances. Moreover, since type II SNe are produced essentially by young stellar populations, they may constitute a more homogeneous sample, than type Ia SNe, with respect to the age of the stellar population. However, it should be noted that they are sgnificantly fainter, and that their study could be more difficult, since they may explode in younger and dustier regions, and this especially holds at cosmological distances, in a general younger environment. On the other side, they are expected to be more abundant per unit volume \citep{Cappellaro2005,Hopkins2006}.

All the current calibrations of SCM basically rely on a sample of type IIP SNe spanned in a range in $z$, for which magnitudes and expansion velocities were available. The major uncertainties are:

\begin{itemize}

\item explosion epoch: unless a very early detection and very good sampling of the phenomenon is available, since the method requires a common reference epoch \citep[day $+50$, but other choices are allowed, see][]{Olivares2010}, an uncertainty in the explosion epoch reflects in a scatter in the derived calibration, especially because the expansion velocity rapidly changes during the first $\approx 30$ days;

\item reddening correction: only for bright SNe explosions in nearby galaxies sound estimates of the local reddening are available, for example \textit{via} the NaI~D. In all the other cases, we must rely on color corrections whose precision is of the order of $0.4$ mag in $(V-I)$ \citep{Olivares2010}. However, this problem has a much lower impact in the NIR bands;

\item velocity of the ejecta: the Fe II $\lambda 5169$ can be difficult to be measured, especially for faint and/or distant SNe, and for the bright SNe the typical uncertainties are of the order of $150$ km s$^{-1}$, that is uncertainties of the order of $2-3 \%$ in the measure (at day $50$). Given the typical slopes of the SCM, of order of $5-6$ in the $I-$band, the final contribution to the error budget is of the order of $10-20 \%$;

\item $K$-corrections: when calibrating the SCM in a given photometric band using SNe at different $z$, it should be noted that the photons collected in the observed band, for a given distant SN, actually come from a different wavelength. This means that we are using photons coming from \textit{intrinsically different} forms of the SCM, with different slopes. Therefore, if $K$-corrections are not applied, the observed relationships are likely to be affected by a larger scatter. Moreover, since the published calibrations are based on SNe spanning different ranges in $z$, this could also affect the coefficients. However, \citet{DAndrea2010} correctly adopted $K$-corrections.

\end{itemize}

\begin{figure}
\includegraphics[width=1\textwidth]{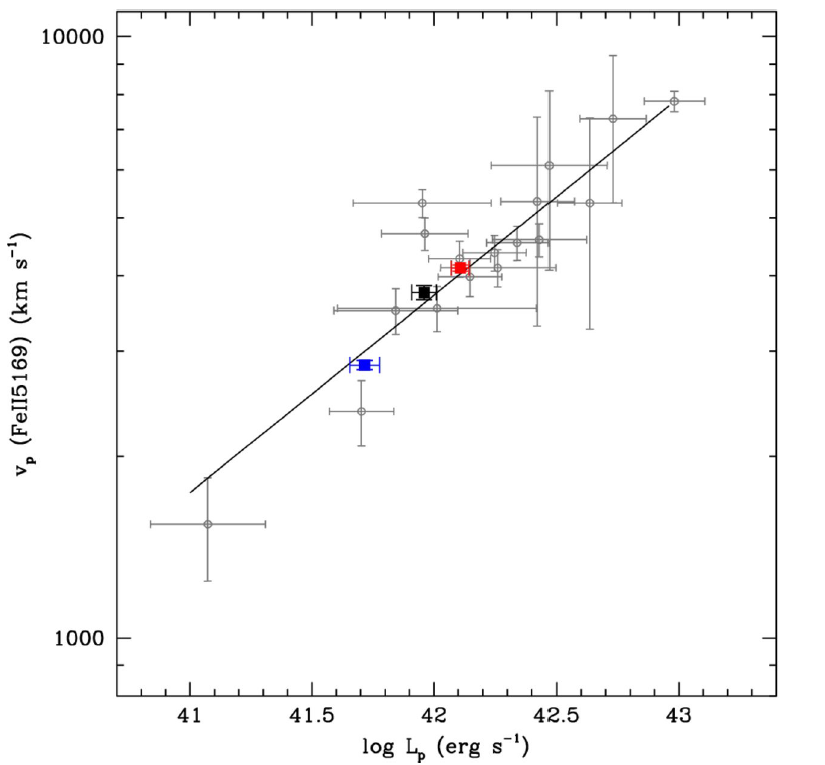}
\caption{Original HP02 plane, with the positions of the homogeneously studied SNe 2012A (blue), 2012aw (red) and 2012ec (black). Plot from \citet{Barbarino2015}, reproduced by kind permission.}
\label{fig:HP02}       
\end{figure}

It should be noted that, when nearby SNe are used, with a tight sampling of their evolution and a homogeneous technique of analysis is employed, the observed scatter of the SCM would be greatly reduced, as suggested in \citet{Barbarino2015}. In Fig.~\ref{fig:HP02} we reproduce their Figure 20, where the position of the SNe 2012A, 2012aw and 2012ec is shown in the original Hamuy \& Pinto plane.

At the present time, we still lack a homogeneous and sound calibration of the SCM based only on the primary distance indicators and, on the other side, we have only a few numbers of host galaxies where type IIP SNe have been exploded and also we detected Cepheids and/or the TRGB. A first step toward this direction was provided by \citet{Hamuy2004} and by \citet{Jang2014}, where they adopted the calibrations of the SCM published by HP02 and \citet{Olivares2010}, respectively, with the distances provided by Cepheids and TRGB, to estimate the Hubble constant. More recently, \citet{Polshaw2015} applied the SCM to SN 2014bc, exploded in the anchor galaxy NGC 4258, for which a geometric maser distance \citep[with an uncertainty of only $3 \%$,][]{Humphreys2013} and a Cepheids distance \citep{Fiorentino2013} is available. They applied almost all the currently available calibrations of the SCM to SN 2014bc, and compared the estimated distances with both the maser and the Cepheids distances. They found some discrepancies between the SCM-based distance moduli and the maser distance modulus, ranging from $-0.38$ mag to $0.31$ mag. To further investigate the scatter among the available calibrations, they applied the SCM to a set of $6$ type IIP SNe occurred in galaxies for which Cepheid distances were available. They obtained a Hubble diagram in the $I$-band with a small scatter ($\sigma_I \sim 0.16$ mag), and the following SCM calibration:

\begin{eqnarray}
I_{50} - A_I &+&  5.665 (\pm 0.487) \log(v_{50}/5000) =  \nonumber \\
 & 5 & \log (H_0D) -2.045 (\pm 0.137)
\end{eqnarray}
where $H_0 = 73.8$ km s$^{-1}$ Mpc$^{-1}$ \citep{riess_2011}, and $D$ is the distance.

This calibration relies on the cosmic distance ladder, even if based on only a few objects. However, the derived SCM is based on Cepheid distances based on \textit{different} calibrations of the Cepheid period-luminosity relations. The differences among the various calibrations are typically of the order of $0.1$ mag \citep[e.g.][and references therein]{Fiorentino2013}. Moreover, also the adopted reddenings came from different sources, even from different calibrations of the same NaI~D feature. To this aspect, we point out that current calibrations of the NaI~D feature in the same galaxy may provide differences up to $\Delta E(B-V) \sim 0.15$ mag \citep[see the discussion in][]{DallOra2014}.

A further development of a calibration of the SCM based on primary distance indicators (Cepheids, TRGB) is highly desirable, with a larger number of calibrators and with a \textit{homogeneous} analysis (i.e. same estimate of the reddening and same Cepheids period-luminosity relations and TRGB calibrations). Moreover, since the SNe calibrated on primary distance indicators occur in the local Universe, they are likely to be deeply investigated. Moreover, the progenitor could be detectable on archive images. This would allow us to fully explore the space of the structural parameters that could affect the SCM. Indeed, very recently SN LSQ13fn \citep{Polshaw2016} was found to break the standardized candle relation. A possible explanation for that could be the low metallicity of the progenitor \citep[$\sim 0.1 Z_\odot$,][]{Polshaw2016}. As a matter of fact, theoretical models \citep{Kasen2009} predict a metallicity dependence, but at the $0.1$ mag level, not as large as observed in the case of SN LSQ13fn (almost $2$ mag). However, a possible explanation could be a combined effect of low-metallicity of the ejecta and a strong circumstellar interactions. Whatever the case, detailed studies of nearby type IIP SNe, spanning a range of masses, metallicities and environments are of extreme importance to fully characterize the SCM.

As a final point we note that future facilities, such as E-ELT and NGST, will allow us to extend the range of the cosmological type IIP SNe, on which SCM could be applied, but also the range of \textit{local} SNe, to calibrate the SCM. However, these forthcoming facilities will operate at the NIR wavelengths, therefore making it essential a sound NIR calibration of the SCM.

\section{Tully-Fisher and Faber-Jackson methods}
\label{sec:TF}

Tully-Fisher method is a historically important method based on the empirical relation between the intrinsic luminosity of a spiral galaxy and the emission line width \citep{tully_1977}. This method opened a way to measure distances and proper motions for spiral galaxies and clusters of galaxies. The method was calibrated using Cepheid stars in nearby spiral galaxies, and then could be easily extended up to $\sim 100$ Mpc. For over 20 years this was the most popular method for probing this distance range, but with invent of new methods, particularly SN Ia, it's role in cosmology strongly diminished due to considerable intrinsic scatter. This scatter can be only partially reduced when replacing the optical luminosity with IR emission, molecular hydrogen or CO. Thus, in recent years the Tully-Fisher relation was rather used to probe the mass to luminosity (M/L) variation in galaxies with
known distances, rather than as a distance estimator in itself \citep[e.g.][]{davis_2016}. In principle, once a reference sample for
each morphological type is determined then one could use it to estimate
distances in the usual way, but at this stage it is not yet expected to give reliable results.

Like a Tully-Fisher, Faber-Jackson relation \citep{faber_1976} is an empirical relation between the intrinsic luminosity and the stellar velocity dispersion in the elliptical galaxies. 
However, this method has vary large intrinsic scatter although it was based on some pre-selection of morphological types, and the actual relation between the galaxy kinematics and morphology is rather complex \citep[see e.g.][]{cortese_2016}.

\section{Surface Brightness Fluctuations}
\label{sbf}
In 1988, Tonry \& Schneider developed a new technique for measuring extragalactic
distances based on the spatial luminosity variations in early-type
galaxies \citep{tonry_1988}. The method, known as Surface Brightness Fluctuations (SBF) 
works because the variation in brightness from pixel-to-pixel varies
as the square root of the number of stars per pixel, and thus galaxies at
larger distances will have smaller variations, appearing smoother, than nearby
galaxies \citep[see][ for a recent review of the
SBF technique and distance measurements]{blakeslee_2013}.

Making accurate SBF measurements requires a solid anchor to the 
distance ladder for calibration. In principle, one could determine the SBF
distance calibration based on theoretical modeling of stellar populations;
in practice, it is more common to adopt an empirical calibration, using
Cepheid distances to set the zero point, and the observed variation of
SBF magnitude with color to calibrate stellar population effects.
In addition to measuring distances, SBF has also been used to explore the 
properties of unresolved stellar populations in galaxies with known 
distances, which is valuable in constraining stellar population models 
\citep[see][ for details]{Jensen_2015}.

The ground-based SBF technique was initially calibrated for use at 
optical wavelengths ($I$ and $z$) and in the near-IR ($J$, $H$, 
and $K$-bands). At $I$, the effects of age and metallicity on SBF brightness
are largely degenerate, making it possible to calibrate distances
for a wide variety of early-type galaxies using a single broadband color.
The near-IR bands exhibit more scatter in SBF as a function of galaxy 
color, but the fluctuations themselves are much brighter because the
stellar light in old populations is dominated by red giant branch stars,
which are brighter in the near-IR. 
Even though the IR background is much higher, the brighter fluctuations 
and better seeing (from the ground)
typically make it possible for IR SBF to reach much 
greater distances.

The first ground-based SBF surveys were largely limited to distances of
about 20 Mpc \citep[e.g.,][]{tonry_2001}.
During the last decade, new instruments on the \emph{Hubble Space Telescope} 
(\textit{HST})
have enabled us to achieve unprecedented precision with the SBF method and
push to much larger distances, thanks to their high spatial resolution, point spread function (PSF) stability, low background levels, and relatively wide fields of view.
The Advanced Camera for Surveys (ACS) was used to conduct 
extensive surveys of the Virgo and Fornax clusters, from which a
calibration of the $z$-band SBF distance method with statistical scatter of 
0.08 mag was achieved, corresponding to 4\% in distance 
\citep{mei_2007,blakeslee_2009}.
This puts SBF on par with the most accurate extragalactic distance indicators,
including Type Ia supernovae (SNe) and Cepheids. 

\citet{Jensen_2015} have recently established a new SBF calibration for
the F110W and F160W filters ($J$ and $H$ bands) of the WFC3/IR
camera, which can routinely measure SBF distances to 80 Mpc in a 
single \emph{HST} orbit. 
The new IR SBF calibration is based on the ACS SBF distances to Virgo 
and Fornax galaxies, and is tied to the Cepheid distance scale.
They find a statistical scatter of 0.1 mag (5\% in distance) per 
galaxy for redder ellipticals, with greater variation in bluer and 
lower-luminosity galaxies.
Comparison with stellar population models implies that redder
ellipticals contain old, metal-rich populations, as expected,
and that bluer dwarf ellipticals contain a wider range of stellar
population ages and lower metallicities, with the youngest populations 
near their centers. IR color gradients appear to be closely related to
age, so IR SBF distance measurements are best limited to the reddest
and oldest high mass elliptical galaxies.

A team of astronomers is now using WFC3/IR (PI J. Blakeslee) to measure IR SBF
distances to a sample of 34 high-mass early-type galaxies in the 
MASSIVE survey \citep{ma_2014}. 
The goal of the MASSIVE survey is to better understand
the structure and dynamics of the 100 most massive galaxies within
${\sim}100$ Mpc using a wide array of imaging and spectroscopic
techniques. Of particular interest is measuring the masses of the
central supermassive black holes in these galaxies, for which
accurate distances are necessary. The IR SBF distances will also
remove peculiar velocity errors and better constrain cosmic flows
within 100 Mpc.

The power of SBF as a tool for cosmology is now being established
with a new \emph{HST} program to measure IR SBF distances to a
collection of early-type supernova host galaxies (PI P. Milne).
The goal of this project is to reduce systematic uncertainties in
Type Ia SNe luminosities and explore possible environmental 
effects on the brightnesses of Ia SNe that are typically
calibrated locally using Cepheids in late-type spirals, but are
more often observed in early-type galaxies at high redshift.
IR SBF is the only method that reaches large enough distances to
observe the host galaxies and measure their distances with the 
requisite precision.

Accomplishing the goals of these projects relies on efficient, 
high-accuracy distance measurement that is currently only possible 
with \emph{HST} resolution, and can be accomplished most efficiently 
with WFC3/IR.
The future of IR SBF is not limited to \emph{HST}, however. 
New AO systems on large telescopes such as the multi-conjugate AO 
system ``GeMS'' on the Gemini-South telescope provide a highly
stable and uniform PSF over a wide ($\sim2$ arcmin$^2$) field of view.
Initial GeMS observations of three galaxies have 
demonstrated that the SBF signal can be measured with high 
fidelity in modest exposure times out to 100 Mpc using 
${\sim}0.08$ arcsec FWHM $K$-band images. 

There are plans to continue to develop the
AO SBF techniques with the expectation that the next generation
of large telescope (e.g., TMT, GMT, and E-ELT) with wide-field
AO systems will make reliable IR SBF distance measurements out 
to several hundred Mpc. 
The James Webb Space Telescope also has great potential to 
push the IR SBF technique to distances of perhaps 500 Mpc. 

\section{Active Galactic Nuclei}
\label{sec:AGN}

Active Galactic Nuclei (AGN) are not the objects frequently discussed in the context of distance
measurements. However, an excellent example of the water maser shows their current importance, and 
the continuously increasing AGN samples and our understanding of those objects open new possibilities.

\subsection{Water masers}
\label{sec:H2O}

The discovery of a water maser at 22.23508 GHz in the Seyfert II galaxy NGC 4258 \citep{miyoshi_1995,herrnstein_2005} offers an unprecedented possibility to measure the distance to this galaxy directly through geometrical methods, without any need for intermediate steps, and with high accuracy. 

This coherent emission forms due to collisional excitation \citep[for emission mechanism, see e.g.][]{elitzur_1992,lo_2005}, the emitted frequency corresponds to the transition between rotational energy levels in the water molecule. The emission may form in the medium with the temperature about 300 - 600 K, and particle number densities $10^8 - 10^{11}$ cm$^{-3}$. Its narrow band emission traces precisely the dynamics of the material through the line shifts. 

In general, water maser emission can form in molecular clouds, comets, planetary atmospheres, stellar atmospheres, and in the case of distant galaxies can be mostly found in starburst regions. The maser emission in NGC 4258 is different, it originates in a Keplerian slightly warped disk surrounding the central black hole and it is strong enough that it can be mapped at sub-milli-arc-second resolution by Very Long Baseline Interferometry (VLBI), providing a powerful tool to probe spatial and kinematic distribution of the molecular gas at distances below 1 pc from the central black hole. The resolved image of the emission allows for the measurement of the spacial distribution of the emission, and the wavelength shifts give the velocity, and it was shown already by \citet{miyoshi_1995} that the measured radial profile corresponds to the Keplerian motion. Further measurement of the proper motion, or acceleration, thus opens a way to the measurement of the geometrical distance. Subsequent long-term monitoring of this source \citep{argon_2007,humphreys_2013} allowed to obtain the distance to NGC 4258 with the accurracy of 3 \%. 
New analysis of the same data by \citet{riess_2016} gave the distance measurement to this source:
\begin{equation}
D(NGC~4258) = 7.54 \pm 0.17 {\rm (random)} \pm 0.10 {\rm (systematic)~Mpc,}
\end{equation}
or, equivalently, magnitude distance of $29.39 \pm 0.06$ mag. Here the statistical error is the value coming directly from the Monte Carlo Markov Chain code, used for fitting the parameters, and the statistical error comes from the same code run for various initial values.
This result, combined with a large number of Cepheid stars discovered in NGC 4258 \citep{fausnaugh_2015} and in several hosts of SN Ia, opened a way to determine the local value of the Hubble constant $H_0 = 72.25 \pm 2.38$ km s$^{-1}$ Mpc$^{-1}$ with 2.4\%  accuracy \citep{riess_2016}. 

The water maser in NGC 4258 is not the only one detected in an active galaxy, and not even the first one. The first water maser was detected in 1979 in NGC 4945 \citep{dosSantos_1979}, and a number of water masers are known till now, although none of those sources have such unprecedented data quality from the point of view of a distance measurement. However, statistical use of numerous sources \citep[Megamaser Cosmology Project;][]{braatz_2015} is expected to provide a measurement precision of $\sim 5$ \% on the Hubble constant.

\subsection{Continuum reverberation}
\label{sec:cr}

It is widely thought that accretion onto the supermassive black holes in bright AGN takes place via an optically thick, geometry thin accretion disk described by \citet{SS73}.  Each annulus within such a disk will emit like a blackbody with temperature $T$ at radius $R$.  An accretion disk dominated by heating from viscous dissipation which is around a black hole of mass $M$ accreting at a rate $\dot{M}$, has a temperature profile given by: 
\begin{equation}
T(R) = \left( \frac{3 G M \dot{M} }{8 \pi R^3 \sigma} \right)^{1/4},
\end{equation}
when the radius is much greater than the innermost stable circular orbit. The disk's absolute flux is then given by summing up blackbodies, $B_\nu$, over all disk annuli:
\begin{eqnarray}
f_\nu(\lambda)& = &\int B_\nu(T(R),\lambda)\frac{2\pi RdR\cos i}{D^2} \nonumber \\
& = &
 \left(\frac{1200G^2h}{\pi^9 c^2}\right)^{1/3} \frac{\cos i (M\dot{M})^{2/3}
 \nu^{1/3}}{D^2}.
\end{eqnarray}
Therefore, if the temperature profile of the disk can be measured, the observed and expected fluxes can be compared in order to determine the distance.  The key question, then, is how to determine the temperature profile of the disk.  The accretion disk, expected to be a few light days across, is too small to be spatially resolved, and thus indirect techniques must be used.  Reverberation mapping \citep{blandfordmckee82,peterson14} uses light travel time to measure spatial separations within a distant accretion flow. Much in the same way the continuum variability studies of AGN provide a means to probe the accretion disk around the central supermassive black hole and allow measurement of the temperature profile of the disk \citep[for a detailed description see][]{collier99,cackett07}.  High energy X-ray/EUV photons produced close to the compact object irradiate the surrounding gas which reprocesses this into UV/optical continuum, with the hotter inner regions emitting mainly UV photons and seeing variations in the irradiating flux before the cooler outer regions, which emit mainly optical photons.  As the ionizing radiation varies erratically, so do the reprocessed components but with time delays due to light travel time within the system - the light travel time from source to reprocessing site to observer is longer than that on the direct path from source to observer.  These observable delays provide indirect information on the size and structure of the surrounding accretion flows.  This technique is a powerful probe of accretion flows in AGN. 

Thermal radiation from a disk annulus at temperature $T(R)$ emerges with a range of wavelengths, $\lambda \sim hc/kT(R)$. Roughly speaking, each wavelength picks out a different temperature zone and the time delay $\tau = R/c$ measures the corresponding radius. Thus, in this reprocessing scenario the accretion disk reverberates and we expect to observe correlated variability between all continuum bands, with the shortest wavelength lightcurves varying first.

More specifically, the observed delays between different continuum wavelengths depend on the disk's radial temperature distribution $T(R)$, its accretion rate, and the mass of the central black hole. A disk surface with $T \propto R^{-b}$ will reverberate with a delay spectrum $\tau \propto \lambda^{-1/b}$. For the temperature distribution of a steady-state externally irradiated disk, $T(R) \propto R^{-3/4}$ (as given above), thus the wavelength-dependent continuum time delays should follow
\begin{equation}
 \tau = \frac{R}{c} \propto (M\dot{M})^{1/3}T^{-4/3}
 \propto (M\dot{M})^{1/3}\lambda^{4/3}  \, . 
\end{equation}

Accretion disk reverberation therefore allows the measurement of the disk temperature profile, and so can be used to determine the AGN distance \citep{collier99} via:
\begin{equation}
  D = 3.3 \left(\frac{\tau}{\mathrm{days}}\right)\left(
  \frac{\lambda}{10^4 \mathrm{\AA}}\right)^{-3/2}\left(
  \frac{f_{\nu}/\cos i}{\mathrm{Jy}}\right)^{-1/2} \, \mathrm{Mpc}.
\end{equation}
Here, $f_\nu$ must be the accretion disk flux, which can be obtained by taking difference spectra to isolate the \textit{variable} component of AGN light.  

\begin{figure}
\centering
\includegraphics[width=0.7\textwidth]{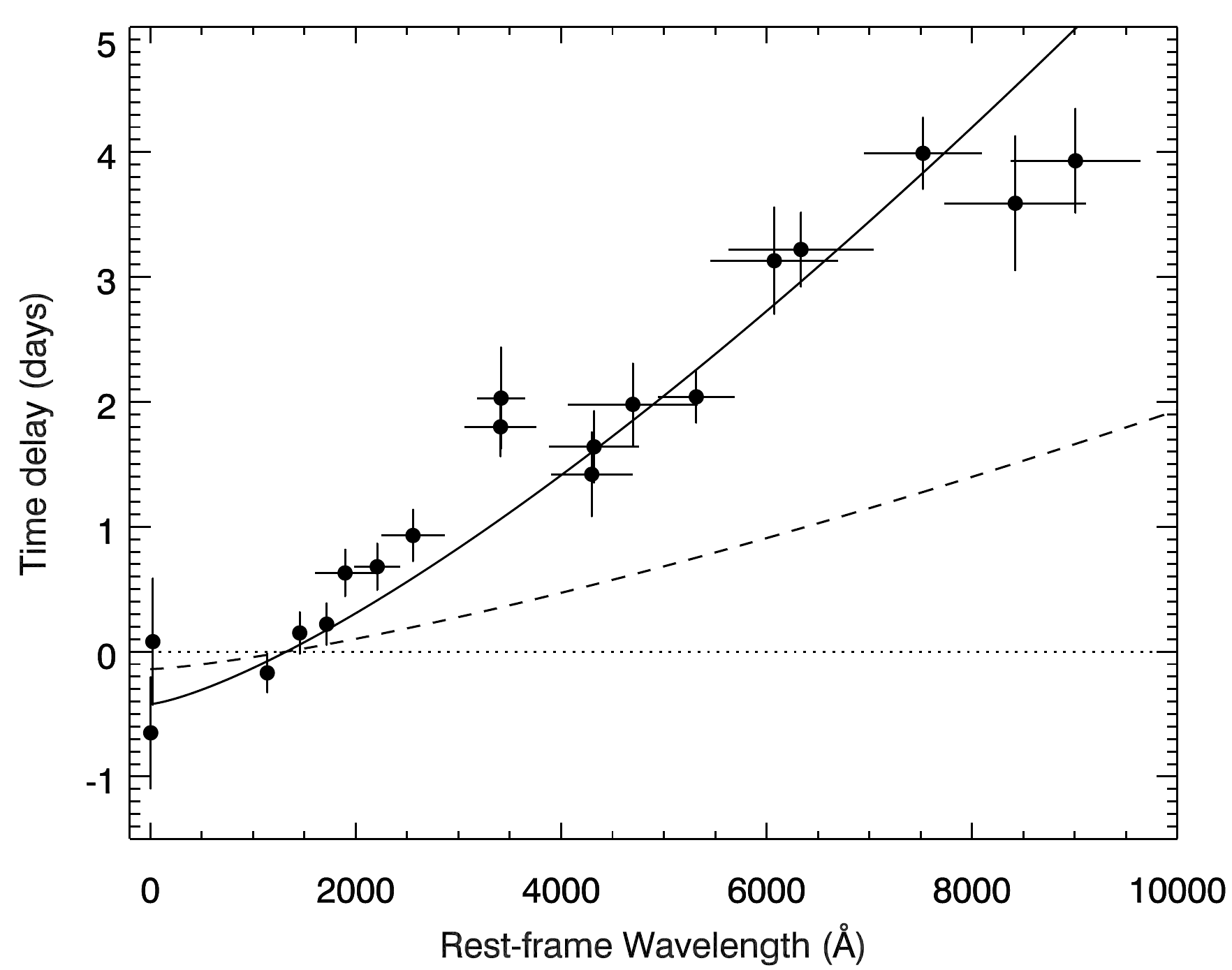}
\caption{Time delay versus rest-wavelength for NGC~5548 \citep[adapted from ][]{fausnaugh16}. The solid line shows the best-fitting $\lambda^{4/3}$ relation, while the dotted line shows the expected lags from a thin disk accreting at 10\% of the Eddington rate.}
\label{fig:ngc5548}
\end{figure}

Over the past 20 years many studies have searched for the expected signatures and found good correlations between different UV and optical wavelength-bands, but the expected interband lags were never of high significance \citep[e.g.,][]{edelson96,wanders97,collier98,collier01,sergeev05}.  Importantly, though, \citet{sergeev05} found that in all the 14 nearby Seyferts in their sample the time-delays are present and show an increase in time delay with increasing wavelength, and the delays increase with greater absolute luminosity of the AGN as predicted by reprocessed disk models.  The first attempt to apply this disk reverberation method to determine an AGN distance yielded an $H_0$ value a factor of 1.66 smaller than any of the currently considered values \citep{collier99}. \citet{cackett07} applied this model to fit the lags and fluxes from all 14 AGN in the Sergeev et al. sample, but again found a value of $H_0$ a factor of 1.6 too small.  One interpretation of this result is that the accretion disks are, on average, a factor of 1.6 larger than predicted by the standard model.

Over the last couple of years, significant progress has been made in better understanding the wavelength-dependent lags.  Much improved sampling cadence and long baseline campaigns have allowed for significant lags from X-rays through to the near-IR to be measured in two objects, NGC 2617 and NGC 5548 \citep{shappee14,mchardy14,edelson15,fausnaugh16}.  Both are consistent with the $\tau \propto \lambda^{4/3}$ relation, however, the lags are a factor of $\sim$3 larger than expected based on the standard disk model \citep{mchardy14,edelson15,fausnaugh16}.  Figure~\ref{fig:ngc5548} shows the wavelength-dependent lags from NGC~5548.

The challenge now lies in understanding the discrepancy with the disk model. Strong emission lines present in broadband filters can act to lengthen the measured continuum lag, since the broad emission line lag is longer than the continuum lag (Chelouche \& Zucker 2013).  However, in NGC~5548 this does not appear to be a large effect \citep{fausnaugh16}.  Furthermore, continuum emission from diffuse gas in the broad line region will also act to lengthen the lag \citep{koristagoad01}, and hints of this effect are present in the NGC 5548 lags, where the u-band (where the Balmer diffuse continuum will peak) lag is an outlier in the otherwise smooth wavelength-dependent lags \citep{edelson15,fausnaugh16}. Alternatively, our understanding of disk accretion is incomplete, and other, more complex geometries occur \citep[e.g.][]{gardnerdone16}.

Future intense monitoring campaigns will help better understand these differences with the standard disk model.  If these discrepancies can be understood, then this method of measuring distances has the potential to be very powerful given the AGN monitoring that will take place in the LSST-era. 


\subsection{BLR reverberation}
\label{sec:BLR}

Intense emission lines, most notably Balmer lines, with kinematic width of order of thousands km/s are the most characteristic features of AGN.
The variability of the emission line intensity has been noticed already by \citet{andrillat_1968}, and the response of the lines to the variable intrinsic continuum opened a way to reverberation mapping of the Broad Line Region \citep[e.g.,][; for a review, see \citealt{peterson_2014,bentz_2015}]{cherepashchuk_1973,gaskell_1986,kaspi_2000}. In the simplest approach, the time delay between the lines and the continuum measures the size of the BLR. Subsequent studies showed that the BLR clouds are predominantly in the Keplerian motion so their orbital velocity, measured through the line width, combined with the orbital radius allows to measure the black hole mass through the virial theorem. This application firmly established the importance of the BLR reverberation for cosmology.

The next important step has been made by \citet{watson_2011} who suggested that BLR reverberation measurement can be also used to determine the distance to the source. AGN are clearly not standard candles, as their intrinsic luminosities span orders of magnitutes, but the reverberation studies showed a tight relation between the delay of the emission line (mosty H$\beta$) and the intrinsic luminosity flux \citep[e.g.,][]{peterson_1999,kaspi_2000}, most frequently measured at 5100 \AA~since the majority of the reverberation studies were done for low redshift  AGN in the optical band. Line delay can be interpreted as a mean/effective radius of the BLR region. Thus this relation, after careful subtraction of the host galaxy reads
\begin{equation}
\log R_{BLR} = K + \alpha \log L_{44},~~~{\rm [light~days]}
\end{equation}
where $L_{44}$ is the monochromatic $\lambda L_{\lambda}$ flux expressed in units of $10^{44}$ erg s$^{-1}$, and the coefficient values from the sample version {\it Clean} given in Tab.~14 of \citet{bentz_2013} are $K = 1.555 \pm 0.024$, $\alpha = 0.542 \pm 0.027$.
The directly measured quantities are the line delay, $\tau$, and the observed flux, $F_{\lambda}$ so the value $\tau/F_{\lambda}^{1/2}$ provides the distance indicator \citep{watson_2011}.

The dispersion in the delay - luminosity relation is only 0.13 dex, as measured by \citet{bentz_2013} in {\it Clean} variant. Thus, with a large number of objects and broader coverage of the redshift range the method offers a very interesting alternative to SNe Ia.

The unique aspect of the AGN Hubble diagram is that while SN distances are observationally restricted now to less than
$z \sim 1.9$ \citep{riess_2001,jones_2013}, AGN diagram can be extended to much larger redshifts thus covering the full range of distances with a single method. The additional advantage of the use of quasars is that those objects, on average, are not strongly affected by extinction (they clean efficiently their environment), and they do not show significant evolution of the metallicity with redshift so in principle no hidden evolutionary bias should be present in their BLR properties across the redshift space.

The number of AGN studied in the context of reverberation mapping is not yet large, about 60 objects were monitored in H$\beta$ line \citep[see][ for a recent compilation]{du_2015,du_2016}, and only a handful of sources have other line delays measured. The most distant quasar with tentative time delay, measured for CIV line, is still S5 0836+71 at $z = 2.172$ studied through 7-year monitoring by \citet{kaspi_2007}. However, many monitoring programs are under way \citep[e.g.,][]{czerny_2013,king_2015,lira_2016}. The use of other lines than H$\beta$ has the advantage that the measured time delay is then considerably shorter which is important for high z quasars, where the expected time delay is long due to large black hole mass as well as to the (1 + z) scaling of the intrinsic timescale to the observed one.

\begin{figure}
\centering
\includegraphics[width=0.45\textwidth]{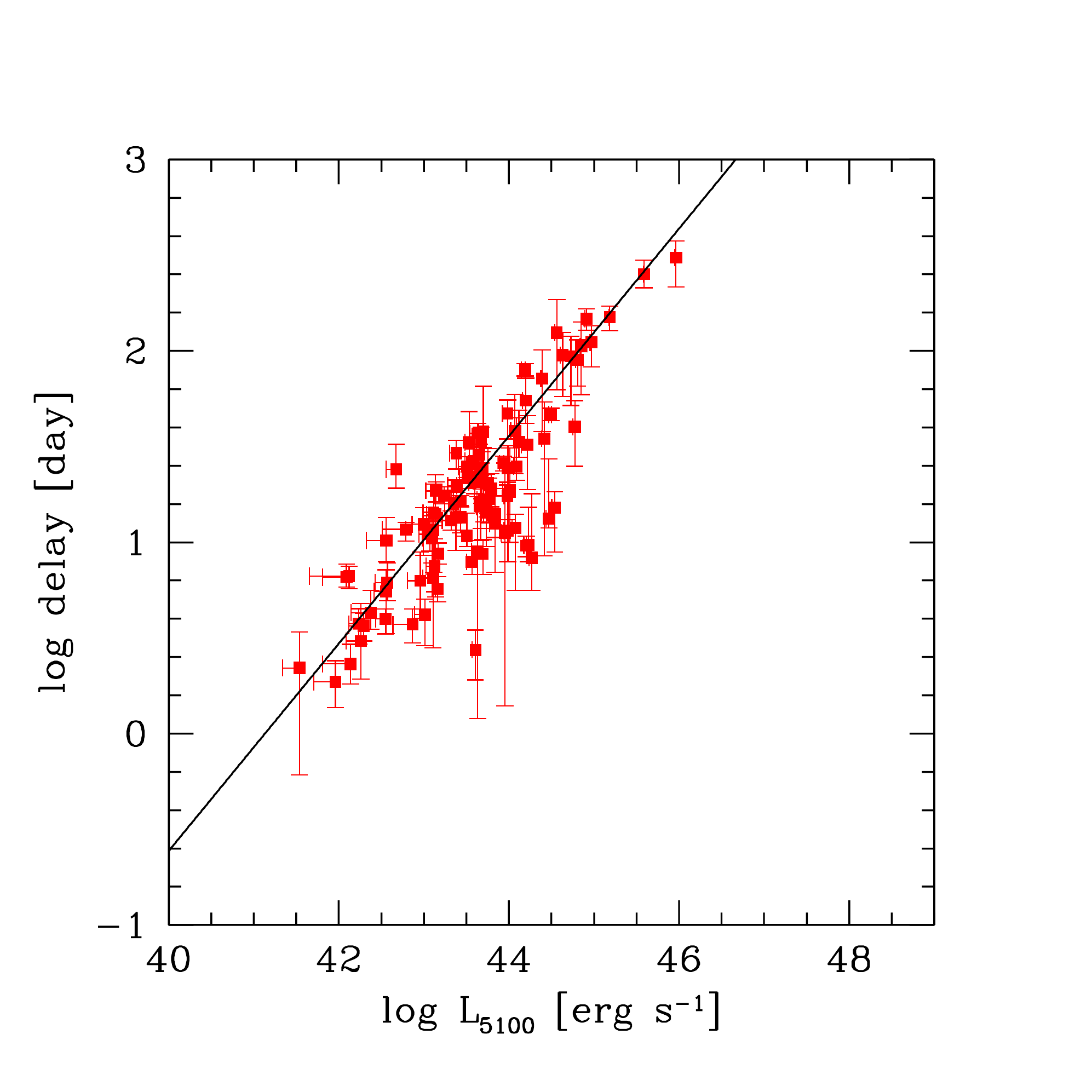}
\includegraphics[width=0.45\textwidth]{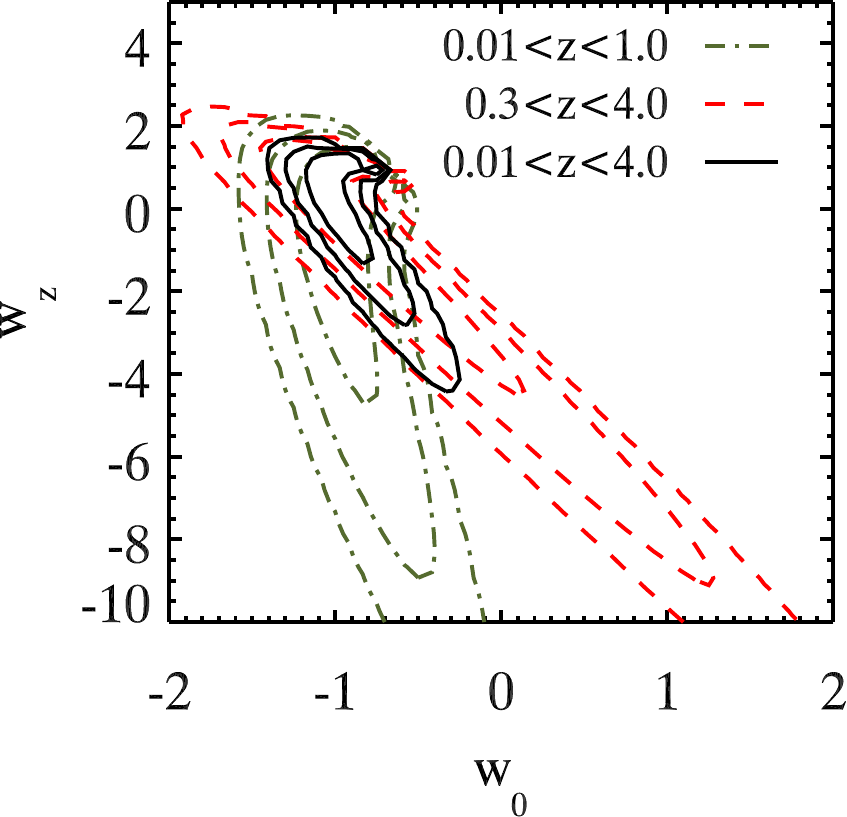}
\caption{Left panel: the time delay of H$\beta$ line as a function of monochromatic luminosity from current reverberation measurements as compiled by \citet{du_2015,du_2016} (red points), and the best fit from \citet{bentz_2013} (see text). Right panel: the expected constraints on dark energy parameters from reverberation-measured 2000 AGN strongly depend on their redshift coverage, and the sources at very low as well as at high redshifts are needed (Fig.~6 from \citealt{king_2014}).}
\label{fig:AGNcosmo}
\end{figure}

Covering the broad range of redshifts with the same probe is very important for distance determination, and for subsequent cosmological constraints. This is best illustrated by \citet{king_2014} where they analyze the expected results from the sample of 2000 AGN. If the dark energy is parametrized as
\begin{equation}
w(z) = w_0 + w_z z/(1 + z)
\end{equation} 
after \citet{chevallier_2001}, then to get the strong constraints AGN must populate the broad redshift range $0.01 < z < 4$, i.e. starting at very low redshifts (see Fig.~\ref{fig:AGNcosmo}). AGN can cover such a broad range while this is a problem for the methods based on gamma-ray bursts (see Sect.~\ref{sec:grb}).

Nearby AGN are currently monitored by several groups \citep[e.g.,][]{ilic_2015,barth_2015,fausnaugh16,bentz_2016a,bentz_2016b,du_2016}. SDSS quasar sample has been recently monitored for six months within the frame of the SDSS-RM project \citep{shen_2016} which allowed to provide preliminary estimate of nine H$\beta$ langs and six Mg II lags for objects with redshifts above 0.3, and this program is being extended now, covering 849 sources in total, with cadence of 12 epochs per year. High redshift quasars do not need such a dense sampling but then the project requires at least 5 years of monitoring (Czerny et al. 2013) to measure lags in higher z sources, and such project is under way within the Oz-DES program \citep{king_2015}. Selected sources (771 in the final sample) will be monitored for six years, with 25 measurements for each quasar, on average. Individual delay measurements are expected for half of them, after carefull subtraction of the Fe II and Fe III emission, host galaxy contamination and correction for the reddening. For fainter sources stack analysis will be performed in a number of luminosity bins. Monitoring of a few $z \sim 1$ quasars with as large telescope as SALT will allow for better disentangling of the Mg II line from the underlying Fe II pseudo-continuum \citep{hryniewicz_2014,modzelewska_2014}.

Narrow-band and broad band reverberation measurements offer an interesting option since they can be done with smaller telescopes and in a more automatic way. In particular, the future Large Synoptic Sky Survey (LSST) will bring 10-year dense coverage of thousands of quasars in six photometric bands. The challenge is in disentangling the line and continuum variability but preliminary studies indicate this is possible \citep[e.g.,][]{chelouche_2014}.

However, before AGN monitoring can reach maturity in cosmological applications, several issues have to be addressed.
First, the AGN Hubble diagram, like SNe Ia, currently requires
calibration to obtain absolute distances. This is basically done by adopting the current value of the Hubble constant for nearby AGN. Direct comparison of AGN distances with another distance indicator is rare. One such example is through the detection of 11 Cepheid stars in a Seyfert 1 galaxy NGC 4395 \citep{thim_2004}.  However, in principle the calibration issue can be avoided if the understanding of the BLR formation progresses. For example, the idea that the BLR forms as a dust-driven failed wind \citep{Czerny_2011} combined with the theory of accretion disk sets the BLR onset at some specific value of $T_{eff}$ provided by the dust sublimation temperature (see Sect.~\ref{sec:cr}). Thus if the hottest dust temperature can be independently measured for a number of AGN this could allow the reverberation method to move to the class of direct methods. Recent comparison of the standard calibration with dust temperature based calibration for a whole AGN sample implies dust temperature 900 K, lower than the hottest dust temperature generally measured in AGN although much higher temperature value was found using the same model from NGC 5548 \citep{galianni_2013}. 

Second aspect is the objective removal of outliers. Some systematic departure of very high Eddington ratio sources from the overall trend has been recently noted \citep{du_2016fun}. If this conclusion is supported in further studies, such super-Eddington sources have to be removed from the general sample or corrected for the departure trend. 

The third problem is related to possible systematic errors when we move towards more distant, brightest quasars.
The extension of the power law dependence between $R_{BLR}$ and the monochromatic flux in a form of a simple power law has not been observationally tested. If the BLR formation is well explained by the failed dust model, expected departures from the linear trend are not strong but if the BLR radius mostly depend on the total ionizing flux then the decrease in ionizing photons with the black hole mass (larger at larger distances) may cause strong nonlinear behavior. The measured monochromatic flux is also affected by the viewing angle, usually unknown. This is not a large problem for nearby AGN since the viewing angles of type 1 AGN are limited to the range between 0 and $\sim 45$ deg due to the presence of the dusty torus, and the lag independence on the viewing angle was demonstrated by \citet{starkey_2016}. However, if the torus opening decreases with redshift and/or with luminosity it may lead to some systematic errors. Recent studies imply some dependence of dust coverage on luminosity \citep[e.g.,][]{ichikawa_2016} but the opposite conclusion has been reached by \citet{mateos_2017}, so more future studies along this line are still needed.

Therefore, the BLR reverberation method has not yet reached maturity. On the other hand, it has a large future potential. Large samples of reverberation-studied sources are coming. Extension of the method to photometric reverberation \citep{haas_2011} opens a way for future use of the Large Synoptic Sky Survey (LSST) data for this purpose, which will bring 10 years of quasar monitoring, with the cadence of 100 observations per year in 6 photometric channels. In the meantime, systematic problems can be studied with increasing sample of individual, well spectroscopically monitored objects. 

\subsection{Extragalactic distances based on dust reverberation of AGNs}
\label{sec:dr}

The inner radius of the dust torus in an active galactic nuclei (AGN)
 is considered to be determined by sublimation of dust
 and is proportional to
 the square-root of the accretion-disk luminosity.
 If its physical size can be measured by reverberation of dust emission,
 then the luminosity distance of an AGN can be obtained.
 The Hubble constant was estimated from the distances based on
 the dust reverberation for local AGNs as $73$ km s$^{-1}$ Mpc$^{-1}$,
 which shows good agreement with its current standard estimates. 

 Near-infrared interferometry recently begins to be able to measure
 the angular scale of the innermost dust torus for brightest AGNs,
 and the angular diameter distance can be obtained
 by comparing the reverberation radius with it.

\subsubsection{Luminosity distance based on the dust reverberation}

\begin{figure}
 \begin{center}
 \includegraphics[width=0.65\textwidth]{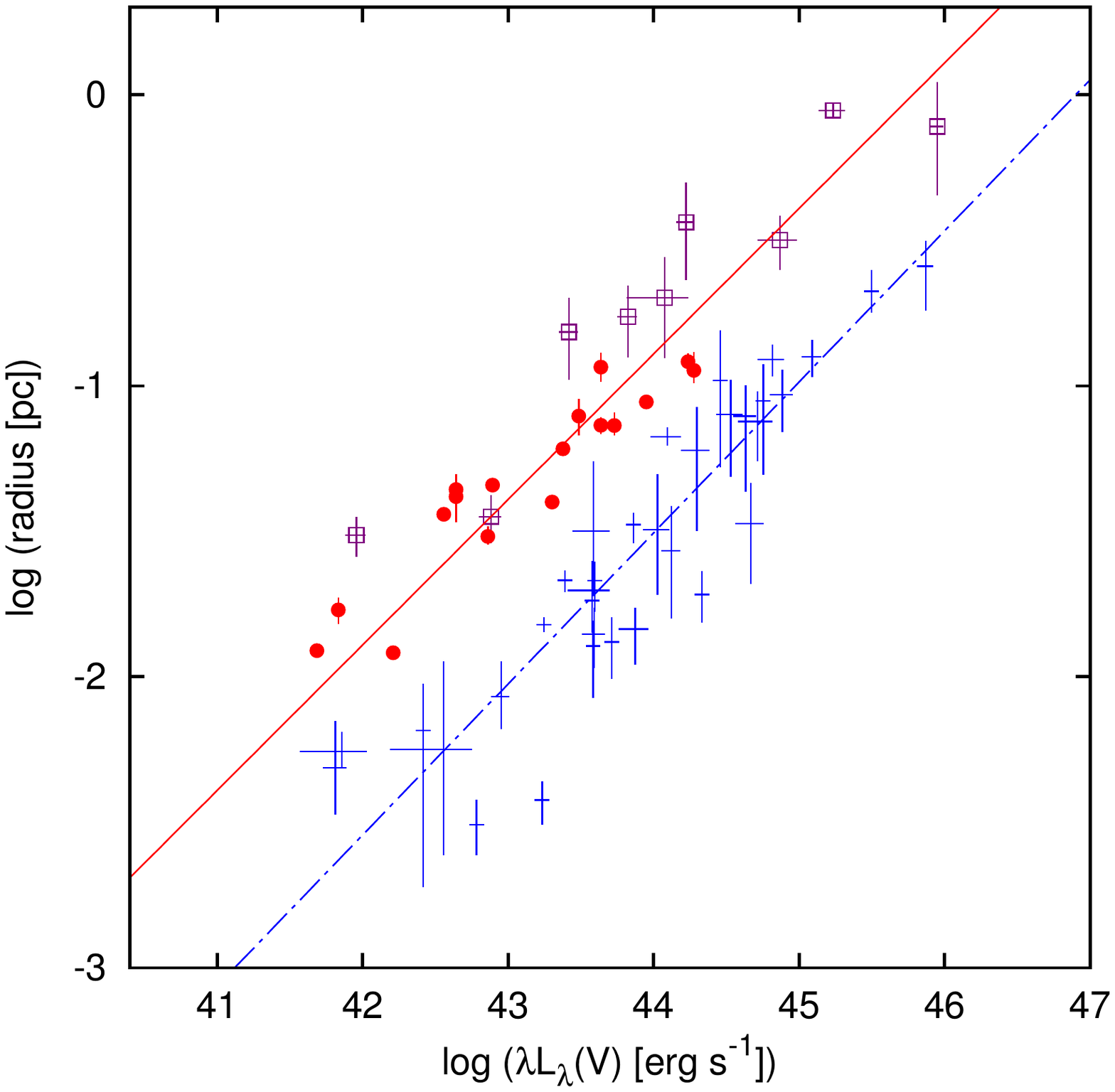}
 \end{center}
 \caption{Radii of the innermost dust torus and the BLR plotted
 against the $V$-band luminosity \citep{Kos14}.
 The red filled circles represent the reverberation radii for the dust torus
 obtained by the MAGNUM project,
 and the red solid line represents the best-fit regression line for them.
 The blue crosses represent the reverberation radii for the BLR
 and the blue dot-dashed line represents the best-fit regression line
 for them \citep{Ben09}.
 The purple open squares represent the interferometric radii in near-infrared
 for the dust torus \citep{Kis11,Wei12}.
 }
\label{fig:1}       
\end{figure}

 Many observations have indicated that an obscuring dust torus
 surrounds an accretion disk and broad emission-line region (BLR)
 in the center of an active galactic nucleus \citep{Ant93}.
 Since dust is sublimated in the vicinity of the accretion disk
 by absorbing its strong UV-optical continuum emission,
 the inner radius of the dust torus is considered to be determined
 by sublimation of dust.
 The dust sublimation radius $r_{\rm d}$ can be derived from
 the radiation-equilibrium equation for the dust grain,
\begin{equation}
\pi a^2\int Q_{\nu }\frac{L_{\nu }^{\rm AD}}{4\pi r_{\rm d}^2}d\nu =4\pi a^2\int Q_{\nu }\pi B_{\nu}(T_{\rm sub})d\nu
\end{equation}
 where $T_{\rm sub}$ is the dust sublimation temperature,
 $a$ is the dust grain size, $Q_{\nu }$ is the absorption efficiency of dust,
 and $B_{\nu}$ is the Planck function \citep{Bar87}.
 The parameters $T_{\rm sub}$ and $Q_{\nu }$ at the innermost region
 of the dust torus would be common in AGNs,
 because only the dust grains with highest sublimation temperature
 can survive there and those parameters are determined
 by the properties of such dust grains.
 Therefore, the inner radius of the dust torus is expected to be
 proportional to the square-root of the accretion-disk luminosity.
 Inversely, the absolute luminosity can be estimated
 once the inner radius of the dust torus is obtained observationally,
 which could be used as a distance indicator for AGNs.

 The dust reverberation enables us to obtain
 the inner radius of the dust torus
 by measuring the lag between the flux variation
 of the UV-optical continuum emission from the accretion disk
 and that of the near-infrared thermal emission from the dust torus.
 A possible application of the dust reverberation of AGNs
 to the cosmological distance measurement
 was proposed at the end of the 20th century \citep{Kob98,Okn99,Yos02},
 and the group of the University of Tokyo
 started the Multicolor Active Galactic NUclei Monitoring (MAGNUM) project
 (PI: Yuzuru Yoshii) in 1995
 to establish the distance indicator based on the dust reverberation of AGNs
 and to constrain the cosmological parameters \citep{Yos02}.

\begin{figure}
 \begin{center}
  \includegraphics[width=0.65\textwidth]{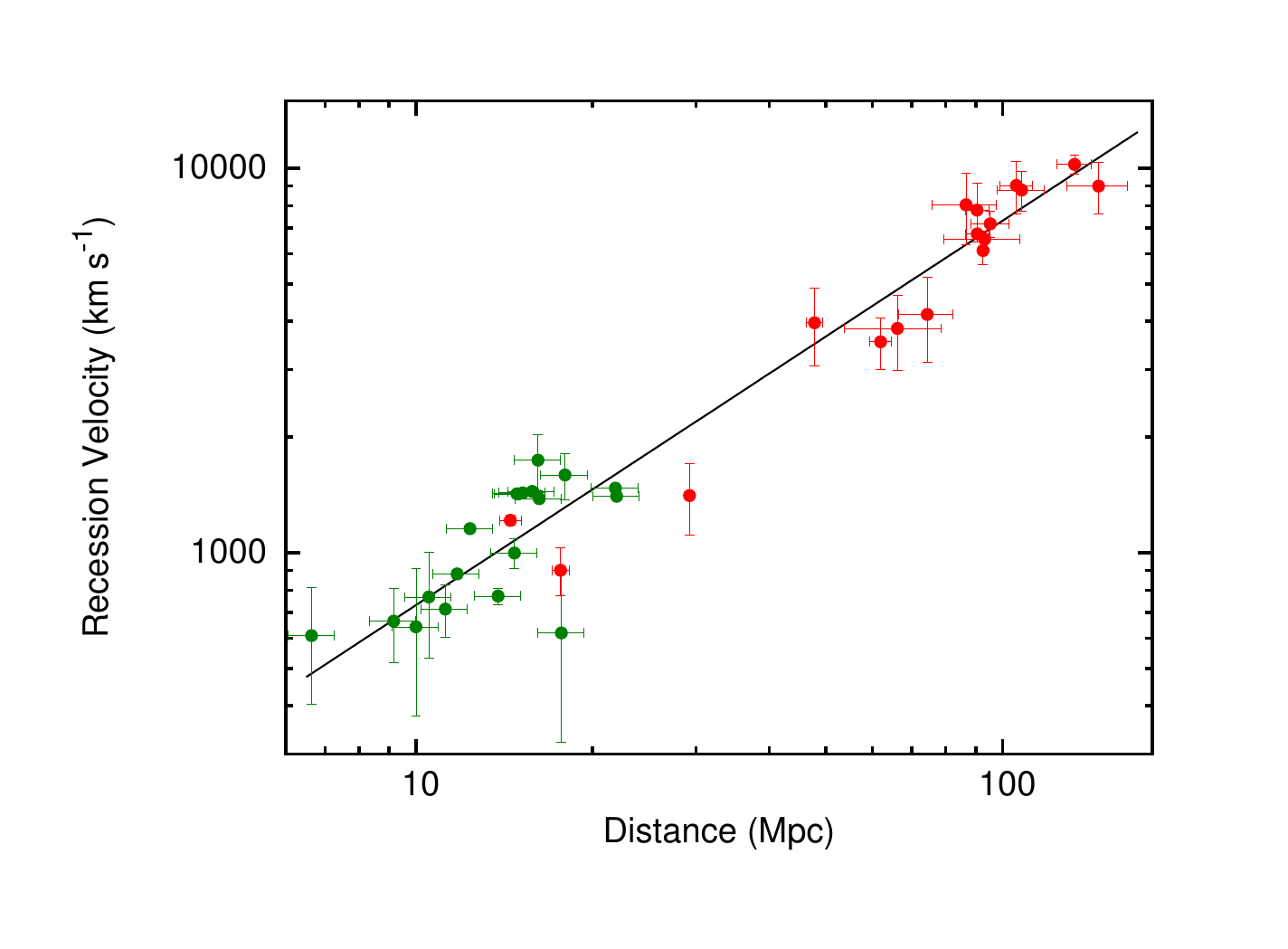}
 \end{center}
 \caption{Hubble diagram for galaxies with
 the dust reverberation distance (red filled circles)
 and the Cepheid-variable distance (green filled circles) from \citet{Yos14}.
 The solid line represents the best-fit regression line
 for the data based on the dust reverberation distance.
 }
\label{fig:2}       
\end{figure}

 Figure~\ref{fig:1} shows the radius-luminosity relation for the dust torus
 obtained by the largest systematic dust reverberation survey
 for 17 local Seyfert galaxies performed by the MAGNUM project \citep{Kos14}.
 The inner radius of the dust torus clearly correlates with
 the optical luminosity of AGNs as theoretically expected,
 which demonstrates the feasibility of the luminosity distance indicator
 based on the dust reverberation.
 Then, the dust sublimation model at the innermost dust torus
 was built to obtain the distances for the 17 AGNs
 without any distance ladder,
 and the distances were compared with the recession velocities
 to estimate the Hubble constant \citep{Yos14}.
 Figure~\ref{fig:2} shows the Hubble diagram.
 The Hubble constant was estimated approximately as
 $H_0=73$ km s$^{-1}$ Mpc$^{-1}$,
 which shows good agreement with its current standard estimates.
 In addition, the distance calibration based on the dust
 sublimation model in \citet{Yos14} was consistent with
 that obtained from the distances of the SNe Ia occurred in
 the AGN host galaxies \citep{Kos17}.
 These results indicate that the distance indicator
 based on the dust reverberation is a promising new tool
 for investigating the expanding Universe.

\subsubsection{Trigonometric parallax for the dust torus}
\label{section:2}
 When both the physical size and the angular size of a certain structure
 in an AGN can be obtained, the angular diameter distance
 can be estimated by trigonometry.
 Although usual imaging observations cannot resolve
 the inner structures of AGNs whose physical sizes are obtained
 by reverberation mapping,
 their angular size might be able to be obtained by
 future long-baseline interferometers in optical and near-infrared wavelengths.
 This trigonometric parallax for AGNs was first proposed
 for the BLR \citep{Elv02},
 but actually, it has not been applied to any AGNs yet.

 Since the dust torus is more extended than the BLR,
 the trigonometric parallax for the dust torus would be more feasible.
 Indeed, recent advancement of near-infrared interferometry
 makes it possible to measure the angular scale of the innermost dust torus
 for some brightest AGNs \citep{Kis11,Wei12},
 and the angular diameter distance of NGC 4151 was estimated
 by comparing the dust reverberation radius with
 the interferometric angular scale in near-infrared \citep{Hon14}.
 However, as shown in Figure~\ref{fig:1},
 the interferometric radius was found to be systematically larger
 than the reverberation radius by about a factor of two \citep{Kos14},
 whereas it is not the case for NGC 4151.
 Further studies on the structure of the dust torus
 are necessary to understand this systematic difference
 and to extend the use of the dust parallax distance of AGNs.

\subsection{Single epoch methods}

Reverberation methods are time-consuming, and much more efficient alternative is to design the method based just on a single spectrum. There are some options in consideration, but they are still in their infancy.

If  Eddington  ratio  and black hole mass  can  be  derived  from  some  distance-independent  measure  it
would be possible to derive distance-independent quasar luminosities. \citet{sulentic_2014} suggest that quasars radiating close to the Eddington limit show distinct
optical and UV spectral properties that can be recognized in spectra. They propose specific criteria based on the line ratios (Al III$\lambda$1860/Si III]$\lambda$1892 $\geq$ 0.5, and (ii) Si III]$\lambda$1892/C III]$\lambda$1909 $\geq $ 1.0) which lead to a source sample with dispersion of $\sim $ 0.13 dex in the Eddington ratio. Larger sample of such sources may give interesting constraints on cosmological parameters.

\citet{wang_2014} proposed instead to use the basic prediction of the slim accretion disk theory \citep{abramowicz_1988}: the saturation of the luminosity for high Eddington ratio sources.
In slim disks the radiative efficiency drops with the Eddington ratio, so effectively the rise of the total source luminosity is only logarithmic in accretion rate.  Their preliminary tests of cosmological applications are attractive but the method still requires better analysis of the saturation level as well as efficient method of selecting such AGN in large AGN catalogs. 

\subsection{Continuum shape method}
\label{sec:crm}

Quasars can be considered a reliable cosmological tools in a similar way as for Type Ia supernovae, provided that the quasar sample is large enough and there is a relation able to standardize the quasar emission.
A non-linear relation between the optical and X-ray luminosity in quasars was discovered with the first X-ray surveys in the early '80s \citep{tananbaum_1979,zamorani_1981,avni_1986}, and it has been confirmed since with various samples of a few hundred quasars observed with the main X-ray observatories over a redshift range from 0 to $\sim$6.5 and about four decades in optical luminosity.

The largest samples published so far include (i) a compilation of 367 quasars from different 
optical surveys and observed by {\em ROSAT}, {\em XMM--Newton} and {\em Chandra} (333 from \citealt{steffen_2006} and 34 from \citealt{just_2007}); (ii) 350 sources obtained from 
the {\it Sloan Digital Sky Survey} 5$^{\rm th}$ quasar catalog with available {\em XMM--Newton} observations \citep{young_2010}; 
(iii) 545 objects from the XMM--COSMOS survey \citep{lusso_2010}; and (iv) 200 quasars with UV and X-ray 
observations from the {\em Swift} observatory archive \citep{grupe_2010,vagnetti_2010}. 
In all these works the $\Lx$--$\luv$ relation is parametrized as a linear dependence between 
the logarithm of the monochromatic luminosity at 2500~\AA\ ($\luv$) and the $\alphaox$ 
parameter, defined as the slope of a power law connecting the monochromatic luminosity at 
2~keV ($\Lx$), and $\luv$: $\alphaox=0.384\times\log(\Lx/\luv)$. 
This relation implies that quasars more luminous in the optical are relatively less luminous 
in the X-rays. For instance, the increase by a factor of 10 in $\luv$ implies an increase by only a factor of $\sim$4 in the X-ray luminosity. 
When expressed as a relation between X-ray and UV luminosities, the $\alphaox$--$\luv$ relation becomes $\log \Lx=\beta+\alpha\log \luv$. 
All the works cited above provide consistent values (within the uncertainties) of both normalization and slope, $\beta\sim8-9$ and $\alpha\sim0.6$, with an observed dispersion that ranges between $\delta\sim0.35$--0.40 dex. Luminosities are derived from fluxes through a luminosity distance ($D_{\rm L}$) calculated assuming a standard $\Lambda$CDM model with the best estimates of the cosmological parameters $\Omega_{\rm M}$ and $\Omega_\Lambda$ at the time of the publications.

The potential use of this relation as a cosmological probe is obvious. The  observed X-ray flux is a function of the 
observed optical flux, the redshift, and the parameters of the adopted cosmological model as
\begin{equation} 
\log(\fx)=\Phi(\fuv,D_{\rm L})=\beta'+\alpha\log(\fuv)+2(\alpha-1)\log (D_{\rm L}),
\end{equation}
where $\beta'=\beta + (\alpha - 1)\log(4\pi)$. This relation can be then fitted to a set of optical and X-ray observations of quasars in order to estimate the cosmological parameters. However, none of the samples published so far had, on their own, the size and/or homogeneity to provide useful constraints for cosmological applications. 

\begin{figure}[t]
\sidecaption
\includegraphics[scale=.45]{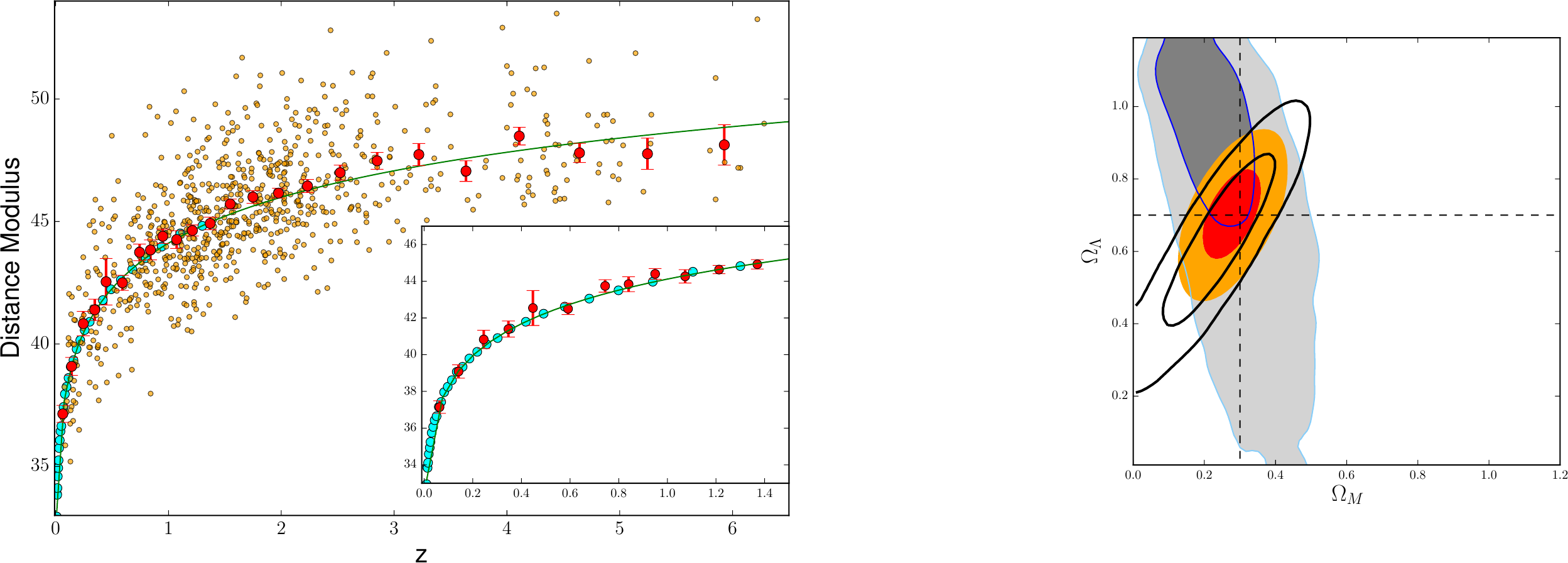}
%
%
\caption{{\it Left}: Hubble diagram for the quasar sample ($\sim$800 sources, orange points), and supernovae (cyan points) from Betoule et al. (2014). The large red points are quasar averages in narrow redshift bins. 
The inner box shows a zoom of the $z=0$--1.5 range, in order to better visualize the match 
between the supernovae and quasar samples. The continuous line is obtained from a joint fit of the 
two samples assuming a standard $\Lambda$CDM cosmological model. 
  {\it Right}: 68\% and 95\% contours for $\Omega_{\rm M}$ and $\Omega_\Lambda$ as derived from 
the Hubble diagram of quasars (grey, Risaliti \& Lusso 2015), from the Type Ia supernovae sample 
(empty black, Betoule et al. 2014), and from a quasar--supernovae joint fit (orange, red). It is the 
first time that this measurement has been done using quasars as cosmological distance probes. Adapted from Risaliti \& Lusso 2015.}
\label{fig:hubblefagiolo}       
\end{figure}

Recently, the relation above has been employed to build a Hubble diagram for quasars to obtain an independent measure of cosmological parameters. \citet{risaliti_2015} have shown that 
it is possible to build a sample large enough to estimate the cosmological 
parameters and to test the $\Lambda$CDM model over the whole redshift range $z=0$--6.5 
(back to an age of the Universe of only $\sim$0.8 Gyr). By using a  
dataset of $\sim$800 quasars from the literature (basically all the ones 
listed at the beginning of this Section) with optical and X-ray data, \citet{risaliti_2015} 
have demonstrated that  the $\fx-\fuv$ relation does not exhibit any dependence with 
redshift, which is a necessary condition to use it as a distance estimator. 
It has been therefore possible to build, for the first time, a {\em quasar Hubble diagram} 
in excellent agreement with that of Type Ia supernovae over the common redshift range 
($z\leq1.4$), but extending it to much higher distances. Figure~\ref{fig:hubblefagiolo} (left panel) 
shows the Hubble diagram for quasars (small black points, while red points are averages 
in narrow redshift bins for visual purpose only), compared to the one of Type Ia supernovae samples (cyan 
points, from \citealt{suzuki_2012}). 
Interestingly, since the same physical quantity (i.e. the distance modulus) is involved, 
this analysis based on the distance modulus--redshift relation allows us to directly merge 
quasar and supernova data on the same diagram, where both are used as {\it standardized candles}.  
This method offers unique access to a region of the distance modulus--redshift plane 
where it is extremely unlikely (if not impossible) to observe any Type Ia supernovae. 

%


The results for the determination of the cosmological parameters $\Omega_{\rm M}$ and $\Omega_\Lambda$ 
are shown in the right panel of Fig.~\ref{fig:hubblefagiolo}. The 68\% and 95\% contours for 
$\Omega_{\rm M}$ and $\Omega_\Lambda$, assuming a standard $\Lambda$CDM model, are derived from the 
analysis of the Hubble diagram of quasars only (grey contours), from the supernovae of the Union 2.1 sample 
(empty black contours; \citealt{suzuki_2012}), and from a quasars--supernovae joint fit (orange/red 
contours). The present data return $\Omega_{\rm M}=0.22^{+0.08}_{-0.10}$ and 
$\Omega_\Lambda=0.92^{+0.18}_{-0.30}$ ($\Omega_{\rm M}=0.28\pm0.04$ and $\Omega_\Lambda=0.73\pm0.08$ from a joint fit).
The contour plot relative to the joint sample is not a statistical intersection of the two individual contours, but the result of a simultaneous fit of the distance modulus--redshift relation for the combined sample quasar/supernovae. This also provides a significantly improved measurement of $\Omega_{\rm M}$ and $\Omega_{\Lambda}$ with pure distance indicators. When all the available cosmological indicators (CMB, BAO, weak lensing, supernovae) are considered, the sample of \citet{risaliti_2015} does not significantly improve the constraints on the cosmological parameters. 
The main limitation of this work is its small predictive power, mainly due to the still large 
dispersion of the $\log \fx=\beta+\alpha\log \fuv$ relation ($\sim$0.30--0.35 dex). For comparison, 
the uncertainty in the distance of a supernova at low redshift is about 0.05 in the same units. This means 
that, in order to obtain the same constraints on a cosmological fit, roughly $(0.35/0.05)^2\sim50$ 
quasars per single supernova are needed. 

The  $\Lx$--$\luv$ relation was recently analyzed on a sample of $>2,600$ SDSS AGN, with X-ray data from XMM--Newton with the aim to understand the origin of the observed dispersion, and to evaluate the true intrinsic dispersion in this relation \citep{lusso_2016}. This work shows that, even though the observed dispersion in the main quasar sample is $\sim$0.45 dex, most of this scatter is caused by the combination of X-ray variability, poor optical/UV and X-ray data quality, and the contamination from the host galaxy and/or AGN with red continua. Once all the contaminations above are taken into account, the real physical dispersion is found to be less than 0.21 dex over roughly 4 decades in luminosity, indicating a tight empirical coupling between disc (UV) and corona (X-rays), which puts the determination of distances based on this relation on a sounder physical grounds.

Although exploratory, the analysis presented in \citet{risaliti_2015} already provides important results: the validation of the UV to X-ray relation and the extension of the Hubble Diagram up to $z\sim6$. The constraints on the cosmological parameters are still loose compared with other methods, and the improvement on the measurement errors obtained by combining the quasar findings with those from supernovae, CMB, BAOs, and clusters is not significant. This work is mainly intended as a demonstration of the method, while most of its potential is still to be exploited in future work, with both available and forthcoming data (e.g. {\it eROSITA}, {\it Euclid}, {\it Athena}) that will significantly increase the number of quasars at $z>2$ where the intrinsic power of this novel technique lies.

\subsection{X-ray excess variance method}
\label{sec:variance}

Another method based on a combination of X-ray observations has been proposed by \citet{laFranca_2014}. The method requires a single optical spectrum (determination of FWHM of one of the strong AGN lines, preferentially H$\beta$) and determination of the X-ray excess variance. The basic idea is simple, it relies on a combination of two methods of the black hole mass determination and subsequent elimination of the mass. The first method, derived from the reverberation campaigns \citep[e.g.,][]{kaspi_2000,peterson_2004} shows that the black hole mass can be determined from a single epoch spectrum, from the monochromatic absolute luminosity (a proxy for the BLR size, see Sect.~\ref{sec:BLR}) and the emission line width. The second method is based on the simple relation between the black hole mass and the high frequency tail of the X-ray power spectrum, or, more conveniently, the X-ray excess variance \citep[e.g.,][]{nikolajuk_2004,gierlinski_2008}. Combining the two and eliminating black hole mass one can determine the absolute luminosity from the X-ray excess variance and the FWHM. The difficulty lies in the high quality measurement of the X-ray excess variance: La Franca et al. argue that the launch of Athena X-ray Observatory (expected in 2028) will provide interesting number of measurements, and they expect more from the future generation X-ray telescopes with larger detector area, but this is not coming soon. 
 
\section{Galaxy cluster distance via X-ray and Sunyaev-Zel'dovich effect}
\label{sec:sz}

Galaxy clusters are the largest virialized astronomical structures in the Universe and observations of their physical properties can provide important cosmological information. An important phenomenon occurring in galaxy clusters is the
Sunyaev-Zel'dovich effect \citep{sunyaev_1972,birkinshaw_1999,carlstrom_2002}, a small distortion of the cosmic microwave background radiation (CMBR) spectrum provoked by the inverse Compton scattering of the CMBR
photons passing through a population of hot electrons. The Sunyaev-Zel'dovich effect (SZE) is
proportional to the electron pressure integrated along the line of sight,
i.e., to the first power of the plasma density. The measured
temperature decrement $\Delta T_{\rm SZ}$ of the CMBR is given by \citep{deFilippis_2005}
\begin{equation}
\label{eq:sze1} \frac{\Delta T_{\rm \scriptscriptstyle SZ}}{T_{\rm
\scriptscriptstyle CMBR}} = f(\nu, T_{\rm e}) \frac{ \sigma_{\rm T}
k_{\rm B} }{m_{\rm e} c^2} \int_{\rm l.o.s.}n_e T_{\rm e} dl, \
\end{equation}
where $T_{\rm e}$ is the temperature of the intra-cluster medium,
$k_{\rm B}$ the Boltzmann constant, $T_{\rm \scriptscriptstyle CMBR}
 =2.728^{\circ}$ K
is the temperature of the CMBR, $\sigma_{\rm T}$ the Thompson cross
section, $m_{\rm e}$ the electron mass and $f(\nu, T_{\rm e})$
accounts for frequency shift and relativistic corrections \citep{itoh_1998}.

Other important physical phenomena occurring in the intra-galaxy
cluster medium are the X-ray emission caused by thermal bremsstrahlung
and line radiation resulting from electron-ion collisions. The X-ray
surface brightness $S_X$ is proportional to the integral along the
line of sight of the square of the electron density. This quantity
may be written as follows
\begin{equation}
S_X = \frac{D_A^2}{4 \pi D_L^2 } \int _{\rm l.o.s.} n_e^2
\Lambda_{eH} dl , \label{eq:sxb0}
\end{equation}
where $\Lambda_{eH}$ is the X-ray cooling function of the intra-cluster
medium (measured in the cluster rest frame), $D_A$ and $D_L$ are the angular diameter and luminosity distances of the galaxy cluster, respectively,  and $n_e$ is the
electron number density. It thus follows that the SZE and X-ray
emission both depend on the properties ($n_e$, $T_e$) of the intra
cluster medium.

As is well known, it is possible to obtain the angular diameter
distance (ADD) of galaxy clusters by their SZE and X-ray surface
brightness observations (SZE/X-ray technique). The calculation begins by constructing a
model for the cluster gas distribution. Assuming, for instance, the
spherical isothermal $\beta$-model such that $n_e$ is given by \citep{cavaliere_1978}
\begin{equation}
n_e({{r}}) =  \left ( 1 + \frac{r^2}{r_c^2} \right )^{-3\beta/2},
\label{eq:iso_beta}
\end{equation}
equations (\ref{eq:sze1}) and (\ref{eq:sxb0})  can be integrated. Here $r_c$ is the core
radius of the galaxy cluster. This $\beta$ model is based on the
hydrostatic equilibrium equation and constant temperature \citep{sarazin_1988}. In this way, we may write for the SZE 
\begin{equation}
\label{eq:sz2} \Delta T_{\rm SZ} = \Delta T_0 \left( 1+
\frac{\theta^2 }{\theta_{c}^2} \right)^{1/2-3\beta/2},
\end{equation}
where { $\theta_{c}=r_c/{{D}}_A$ is the angular core radius and}
$\Delta T_0$ is the central temperature decrement that includes all
physical constants and terms resulting from the line-of-sight
integration. More precisely:
\begin{equation}
\label{eqsze3} \Delta T_0 \equiv T_{\rm \scriptscriptstyle CMBR}
f(\nu, T_{\rm e}) \frac{ \sigma_{\rm T} k_{\rm B} T_{\rm e}}{m_{\rm e}
c^2}n_{e0} \sqrt{\pi} \theta_c D_A g\left(\beta/2\right),
\end{equation}
with
\begin{equation}
g(\alpha)\equiv\frac{\Gamma \left[3\alpha-1/2\right]}{\Gamma \left[3
\alpha\right]}, \label{galfa}
\end{equation}
where $\Gamma(\alpha)$ is the gamma function and the others
constants are the usual physical quantities. For X-ray surface
brightness, we have
\begin{equation}
S_X = S_{X0} \left( 1+ \frac{\theta }{\theta_{c}^2} \right)^{1/2-3
\beta}, \label{eqsxb1}
\end{equation}
where the central surface brightness $S_{X0}$ reads
\begin{equation}
\label{eqsxb2} S_{X0} \equiv \frac{D_A^2 \Lambda_{eH}\
\mu_e/\mu_H}{D_L^2 4 \sqrt{\pi} } n_{e0}^2 \theta_c D_A \ g(\beta).
\end{equation}
Here $\mu$ is the molecular weight given by $\mu_i\equiv
\rho/n_im_p$.

One can solve equations (\ref{eqsze3}) and (\ref{eqsxb2}) for the ADD by eliminating $n_{e0}$ and taking for granted the validity
of cosmic distance duality relation (CDDR), $D_L(1+z)^{-2}D_A^{-1}=\eta=1$. In this case one
obtains
\begin{eqnarray}
\label{eqobl7}
{{D}}_A &= & \left[ \frac{\Delta {T_0}^2}{S_{\rm X0}}
\left( \frac{m_{\rm e} c^2}{k_{\rm B} T_{e0} } \right)^2
\frac{g\left(\beta\right)}{g(\beta/2)^2\ \theta_{\rm c}}
\right] \times \nonumber \\
& & \times \left[ \frac{\Lambda_{eH0}\ \mu_e/\mu_H}{4 \pi^{3/2}f(\nu,T_{\rm
e})^2\ {(T_{\rm \scriptscriptstyle CMBR})}^2 {\sigma_{\rm T}}^2\ (1+z_{\rm
c})^4} \right] \nonumber \\
& 
\end{eqnarray}
where $z_c$ is the galaxy cluster redshift. Recently,  such a technique has been applied for a fairly large number of clusters \citep{reese_2002,deFilippis_2005,bonamente_2006} with systematic and statistical errors around $20\%$ and $13\%$, respectively \citep[see table 3 in][]{bonamente_2006}. The crucial point in the SZE/X-ray technique is the choice of morphology used to describe the galaxy cluster. The standard spherical geometry has been severely questioned, since Chandra and XMM-Newton observations have shown that clusters usually exhibit an elliptical surface brightness \citep{sereno_2006}. The assumed cluster shape can affect considerably the SZE/X-ray distances, and, consequently, the $H_{0}$ estimates and other astrophysical quantities.

\subsection{SZE/X-ray technique applications: the Hubble constant}

The Hubble constant, $H_0$, sets the scale of the size and age of the Universe and its determination from independent methods is still worthwhile to be investigated. Severeal authors used the SZE/X-ray technique to estimate the Hubble constant  \citep[see table in][]{holanda_2012}, however, since the samples used has galaxy clusters in high redshifts (up to $z \approx 0.90$), a cosmological model had to be assumed in analysis, usually a flat $\Lambda$CDM model. Moreover, due to degeneracy on the cosmological parameters, the matter density parameter, $\Omega_M$, was taken as being $\Omega_M=0.3$. An interesting method was adopted by \citet{holanda_2012}. In this case, the authors used a sample of 25 ADD of galaxy clusters in redshift range $0.023 < z < 0.784$ described by an elliptical $\beta$ model \citep{deFilippis_2005} to constrain $H_0$ in dark energy models. In order to avoid the use of priors on the cosmological parameters, a joint analysis involving the ADD, the baryon acoustic oscillations (BAO) and the CMBR Shift Parameter signature was proposed. By taking into account the statistical and systematic errors of the SZE/X-ray technique it was obtained for nonflat $\Lambda$CDM model $H_0 = 74^{+5.0}_{-4.0}$ km/s/Mpc (1$\sigma$) whereas for a flat universe with constant equation of state parameter it was obtained $H_0 = 72^{+5.5}_{-4.0}$ km/s/Mpc (1$\sigma$). These values are in full agreement with the latest local estimate performed by \citet{riess_2016}: $H_0 = 73.24 \pm 1.74$ km/s/Mpc (1$\sigma$).

\subsection{The search for two numbers: $H_0$ and $q_0$}

\begin{figure*}
  \includegraphics[width=0.95\textwidth]{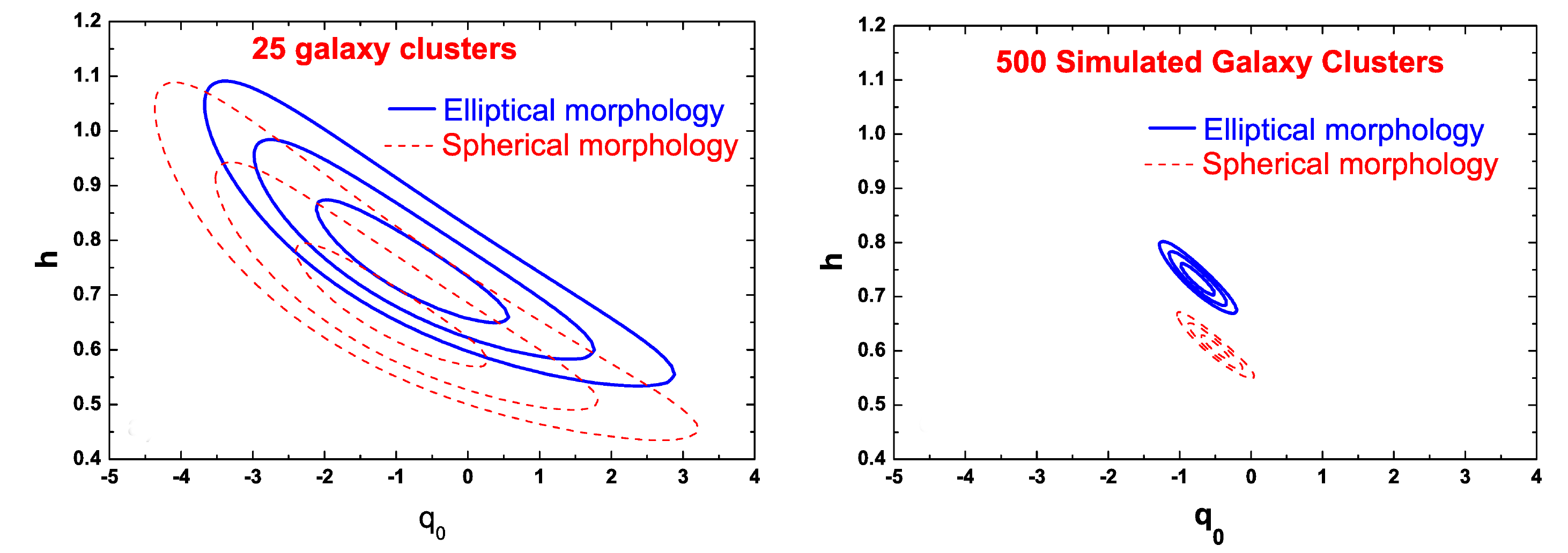}
\caption{ Confidence contours (1, 2 and 3$\sigma$) on the plane ($h - q_0$). The panel left corresponds to the real data.}
\label{fig:sz_clusters}       
\end{figure*}

A kinematic method to access cosmic acceleration and the $H_0$ value  based exclusively on the SZE and X-ray surface brightness data from galaxy clusters also was investigated recently. This is a very important task since until very recently the type Ia supernovae observations was the unique direct
access to the late time accelerating stage of the Universe. By assuming the current observational error distribution of the samples of 25 ADD from \citet{deFilippis_2005}, \citet{holanda_2013} performed Monte Carlo simulations based on a well-behaved parametrization for the deceleration parameter, $q_0$, to generate samples with different characteristics  and study the improvement on the determination of the cosmographic parameters: $q_0$ and $H_0$. As a interesting result it was shown that, even keeping the current statistical observational uncertainty, an increase in the number of data points increases considerably the figure-of merit for the cosmographic plane ($h - q_0$), where $h=H_0/100$ (see figure~\ref{fig:sz_clusters}).

\subsection{The search for new physics}

The SZE/X-ray technique also has been shown to being a powerful tool  to investigate fundamental physics. \citet{uzan_2004}  argued that this technique is strongly dependent on the validity of the CDDR, $D_L(1+z)^{-2}D_A^{-1}=\eta=1$, valid for \textit{all} cosmological models based on Riemannian geometry, being dependent neither on Einstein field equations nor on the nature of matter \citep[][ reprinted as \citealt{etherington_2007}]{etherington_1933}, playing an essential role in modern cosmology. It only requires that source and observer are connected by null geodesics in a Riemannian spacetime and that the number of photons are conserved. However, if $\eta \neq 1$, instead of the real ADD, the measured quantity is $D_A^{\: data}(z)=D_A(z) \; \eta^2(z)$. By using different expressions for $\eta(z)$, several authors have tested the CDDR  by using ADD of galaxy clusters and $D_L$ from type Ia supernovae observations \citep{holanda_2010,holanda_2012b,li_2011,nair_2011,yang_2013,holandaBarros_2016}. No significant deviation was obtained when an elliptical $\beta$ model was used in analysis (for results from different cosmological observations see table I in \citet{holanda_2016}. 

As was showed by \citet{holanda_2016}, this technique depends on the fine structure constant, $\alpha$. If $\alpha$ is a time-dependent quantity, e.g., $\alpha=\alpha_0 \phi(z)$, where $\phi$ is a function of redshift, the current ADD data do not provide the real ADD to the cluster but instead $D_A^{data}(z) = \eta^2(z) \phi(z) D_{A}(z)$. Constraints on a possible variation of $\alpha$ for a class of dilaton runaway models was performed considering the sample of $D_A^{data}(z)$ from \citet{deFilippis_2005}  and estimates of $D_{{A}}(z)$ from type Ia supernovae observations. It was found no significant indication of variation of $\alpha$ with the present data.

Finally, it is very important to stress that the SZE/X-ray technique is independent of any calibrator usually adopted in the determinations of the distance scale. The above results, therefore, highlight the cosmological interest in ADD measurements of galaxy clusters.

\section{Gamma-ray bursts}
\label{sec:grb}

Gamma-rays bursts are the brightest individual sources in the Universe \citep[see e.g.,][ for recent reviews]{hjorth_2011,berger_2014}, therefore it is more than natural to attempt to use them in cosmology. Those intense flashes of the gamma-ray emission lasting from milliseconds to several hours are not easily affected by the medium between an observer and the source, so they can be easily detected at very high redshifts. The record holder is the gamma-ray burst GRB 090429B at $z = 9.4$ \citep{cucchiara_2011}, with photometric redshift determination, or GRB 090429B at $z = 8.2$ \citep{tanvir_2009}, if we limit ourselves to spectroscopically confirmed redshift. Gamma-ray prompted emission is accompanied by afterglow emission at other wavelengths.

However, due to the large burst variety and the critical role of the relativistic boosting of the part of the observed multi-wavelength emission the use of the gamma-ray bursts as distance indicators is rather difficult. The first attempts based on relatively large number of sources was made 
by \citet{schaefer_2007} who successfully constructed the Hubble diagram in a broad redshift range almost up to $z \sim 7$. The method was based on empirical correlations established for sources at $z < 2$ and then extended to higher redshifts.

Several correlations can be used for that purpose \citep[see e.g.,][]{wangLiang_2015}. However, there are two basic problems which are not easy to overcome. First, there is some {\it circularity problem}, as there are no low redshift gamma-ray bursts and the scaling of the relations has to be done for intermediate redshift sources where cosmological effects already become important. Second, the number of gamma-ray bursts with well measured parameters is still low, and it increases only slowly with time. Samples with high quality data and low scatter in the measured quantities contain less than 50 objects so far \citep[e.g.,][]{dainotti_2016}. In addition, the results can be affected by the weak lensing, and the appropriate statistical corrections should be introduced, as discussed by \citet{wangDai_2011}.  

\section{Astronomical distances from gravitational-wave observations}
\label{sec:gw}

Recent direct detections of gravitational waves from merging
binary black hole systems at cosmological distances of $z=0.1-0.2$, 
performed by two Advanced LIGO detectors \citep{GW150914,GW151226,GW170104},
first binary black-hole merger observation with the global network of three
detectors of Advanced LIGO and the Advanced Virgo \citep{GW170814}, as well as
the first Advanced LIGO and Advanced Virgo detection of a nearby (at a distance 
of 40 Mpc) binary neutron-star merger \citep{GW170817}, followed by a short gamma-ray
burst \citep{GW170817GRB} and broad-band electromagnetic emission
observational campaign \citep{GW170817MMA} create unprecedented opportunities
for studying the Universe through a novel, never before explored channel of
spacetime fluctuations.
Gravitational wave astronomy is often compared to
`listening to' rather than `looking at' the skies. By design which is motivated by the
choice of potential sources, the ground-based gravitational wave detectors, 
kilometer-size laser interferometer antennas of 
Advanced LIGO \citep{ALIGO2015} and Advanced Virgo \citep{AdV2015}, 
are sensitive in the range of frequencies similar to the audible range of human
ears - between 10 Hz and a few kHz. As in the case of an ear, a solitary laser
interferometric detector is practically omnidirectional (has a poor angular
resolution), and has no imaging capabilities. It registers a coherent signal emitted
by a bulk movement of large, rapidly-moving masses. Once emitted, gravitational
waves are weakly coupled to the surrounding matter and propagate freely without
scattering. This has to be contrasted with the electromagnetic emission which
originates at the microscopic level, is strongly coupled to the surroundings
and often reprocessed; it carries a reliable information from the last
scattering surface only. It seems therefore that gravitational wave detectors
are the perfect counterpart to the electromagnetic observatories as they may
provide us with information impossible to obtain by other means. 

Very shortly after announcing the general theory of relativity in \citeyear{Einstein1915}, 
Albert Einstein realized that a linearized version
of his equations resembles the wave equation \citep{Einstein1916}.
The solution is interpreted as a short-wavelength, time-varying curvature
deformation propagating with the speed of light on an otherwise slowly-varying,
large-scale curvature background (a gravitational-wave ''ripple'' propagating
through the four-dimensional spacetime); from the point of view of a metric
tensor, it represents a small perturbation of a stationary background metric.
Linear approximation corresponds to the waves propagating in the far-field
limit. By exploiting the gauge freedom of the theory one may show that the
solution has features similar to electromagnetic waves: it is a transverse wave
which may be polarized (has two independent polarizations). Over the next 40
years, during which Einstein changed his mind to argue against their genuineness, 
a controversy persisted over the true nature of gravitational waves. 
Only in the late 50s and early 60s the works of Felix Pirani
(\citeyear{Pirani1956}), Herman Bondi (\citeyear{Bondi1957}), Ivor Robertson and Andrzej
Trautman (\citeyear{RobinsonT1960}) convincingly showed that gravitational waves are
indeed physical phenomena that carry and deposit energy. 

A realistic wave phenomenon (and not, say, a coordinate artifact) must be
capable of transmitting energy from the source to infinity. If the amplitude of an
exemplary isotropic field at a radial distance $r$ from the source is $h(r)$,
then the flux of energy over a spherical surface at $r$ is
$\mathcal{F}(r)\;\propto\;h^2(r)$, and the total emitted power (the luminosity)
is $\mathcal{L}(r)\;\propto\;4\pi r^2 h^2(r)$. Since the energy has to be
conserved, the amplitude $h(r)$ falls like $1/r$, irrespectively of the
multipole character of the source (the lowest radiating multipole in the
gravity theory is the quadrupole distribution, because for an isolated system a
time-changing monopole would correspond to the violation of the mass-energy
conservation, and a time-changing dipole would violate the momentum
conservation law).
 
Gravitational waves are related to the changes in the spacetime distance (the
proper time interval), therefore they cannot be detected by a local measurement
- one has to compare the spacetime positions of remote events \citep{Pirani1956}. 
The detection principle in the case of the laser
interferometric detector is to measure the difference of the relative change in
its perpendicular arms' lengths $L_x$ and $L_y$, $\delta L_x - \delta L_y =
\Delta L/L$, by measuring the interference pattern in the output port located
at the apex of the device. Due to the quadrupolar nature of a gravitational
wave, shortening of one arm corresponds to elongation of the other. This
change of lengths is reflected in varying paths that the laser light has to
cross before the interference. The dimensionless gravitational-wave amplitude
$h=\Delta L/L$ (the ''strain'') is proportional to the amount of outgoing
laser light. The fact that the directly-measurable quantity is the amplitude
$h\;{\propto}\; 1/r$, not the energy of the wave as in the electromagnetic
antenn{\ae}, has a direct consequence for the reach of the observing device:
one-order-of-magnitude sensitivity improvement corresponds to
one-order-of-magnitude growth of distance reach $r$, as opposed to the factor of $\sqrt{10}$
in the electromagnetic observations (consequently, the volume of space grows 
like $r^3$ in case of gravitational-wave observations, encompassing hundreds 
of thousands of galaxies for a distance reach of the order of hundreds of Mpc
(see \citealt{AbbottLRR2016}).    

Among promising sources of gravitational radiation are all asymmetric collapses
and explosions (supernov{\ae}), rotating deformed stars (gravitational-wave
`pulsars' of continuous and transient nature), and tight binary systems of
e.g., neutron stars and black holes. In the following we will focus on the 
binary systems, since their properties make them the analogues to standard
candles of traditional astronomy. The fittingly descriptive term ''standard
siren'' was first used in the work of \citet{HolzH2005} in the context of
gravitational waves from super-massive binary black holes as a target for the
planned spaceborne LISA detector \citep{Danzmann1996}. 
\citet{Dalal2006} were the first one to point out that short GRBs produced by merging binary neutron-star systems would constitute ideal standard siren. The idea of using well-understood signals to infer the distance
and constrain the cosmological parameters, e.g., the Hubble constant, was
however proposed much earlier \citep{Schutz1986,Schutz2002}.  Recently it was
demonstrated in practice for the first time with the observations of a binary
neutron-star merger in both gravitational waves and photons
\citep{GW170817,GW170817H0}.

Magnitudes of the gravitational-wave strain $h$ and the luminosity $\mathcal{L}$ may 
be estimated using dimensional analysis and Newtonian physics. As the waves 
are generated by the accelerated movement of masses and the mass distribution 
should be quadrupolar, one may assume that $h$ is proportional to a second time 
derivative of the quadrupole moment $I_{ij}=\int \rho(\mathbf{x}) x_i x_j d^3x$ for some 
matter distribution $\rho(\mathbf{x})$. For a binary composed of masses $m_1$ and 
$m_2$, orbiting the center of mass at a separation $a$ with the orbital angular 
velocity $\omega$, $h$ is proportional to the system's moment of inertia 
$\mu a^2$ and to $\omega^2$, as well as inversely proportional to the distance, 
$h\,\propto\; \mu a^2 \omega^2/r$, where $\mu=m_1m_2/M$ is the reduced mass,  
and $M=m_1 + m_2$ the total mass. In order to recover the dimensionless $h$, 
the characteristic constants of the problem, $G$ and $c$, are used to obtain  
\begin{equation} 
  h \simeq \frac{G}{c^4}\frac{1}{r}\mu a^2\omega^2 
  = \frac{G^{5/3}}{c^4}\frac{1}{r}\mu M^{2/3}\omega^{2/3}
  \qquad\qquad 
  \left(h_{ij}=\frac{2G}{c^4r}\ddot{I}_{ij}\right), 
\end{equation}  
with the use of Kepler's third law ($GM=a^3\omega^2$) in the second
equation. The expression in brackets represents the strain tensor $h_{ij}$ in
the non-relativistic {\it quadrupole approximation} 
\citep{Einstein1918}. Similarly, the luminosity $\mathcal{L}$ (the rate of energy loss  
in gravitational waves, integrated over a sphere at a distance $r$) should be proportional 
to $h^2r^2$ and some power of $\omega$. From dimensional analysis one has   
\begin{equation} 
  \mathcal{L}  =  \frac{dE_{GW}}{dt} \,\propto\, \frac{G}{c^5} h^2 \omega^2 
  \,\propto\, \frac{G}{c^5} \mu^2 a^4 \omega^6 \nonumber
\end{equation}
\begin{equation}  
  \left(\mathcal{L}  =  \frac{c^3}{16G\pi}\iint\langle\dot{h}_{ij}\dot{h}^{ij}\rangle\;dS 
  = \frac{G}{5c^5}\langle\dddot{I}_{ij}\dddot{I}^{ij}\rangle\right),   
\end{equation}
with the proportionality factor of $32/5$. Again, the expression in brackets refers 
to the quadrupole approximation; the angle brackets denote 
averaging over the orbital period. 
Waves leave the system at the expense of its orbital energy $E_{orb} = - Gm_1m_2/(2a)$. 
Using the time derivative of the Kepler's third law, 
$\dot{a} = -2a\dot{\omega}/\left(3\omega\right)$, one gets the evolution of 
the orbital frequency driven by the gravitational-wave emission: 
\begin{equation}
  \frac{dE_{orb}}{dt} \equiv \frac{Gm_1m_2}{2a^2}\dot{a} = -\frac{dE_{GW}}{dt} 
  \quad\Longrightarrow\quad 
  \dot{\omega} = \frac{96}{5}\frac{\omega^{11/3}}{c^{5}}G^{5/3}\mathcal{M}^{5/3}.  
\end{equation}
The system changes by increasing its orbital frequency; at the same time the
strain amplitude $h$ of emitted waves also grows. This characteristic
frequency-amplitude evolution is called the {\it chirp}, by the similarity to
birds' sounds, and the characteristic function of component masses $\mathcal{M}
= \left(\mu^3M^2\right)^{1/5}$ is called the {\it chirp mass}. Orbital frequency is
related in a straightforward manner to the gravitational-wave frequency
$f_{GW}$: from the geometry of the problem it is evident that the frequency of 
radiation is predominantly at twice the orbital frequency, $f_{GW} =
\omega/\pi$. By combining the equations for $\dot{f}_{GW}$ and $h$, one recovers
the distance to the source $r$. It is a function of the frequency and amplitude
parameters, which are {\it directly} measured by the detector:   
\begin{equation} 
  r = \frac{5}{96\pi^2} \frac{c}{h}\frac{\dot{f}_{GW}}{f^3_{GW}} 
  = 512 \frac{1}{h_{21}}\left(\frac{0.01\;\text{s}}{\tau}\right)
  \left(\frac{100\;\text{Hz}}{f_{GW}}\right)^2\;\text{Mpc}, 
\end{equation}
where $h_{21}$ denotes the strain in the units of $10^{-21}$ and $\tau =
f_{GW}/\dot{f}_{GW}$ denotes the rate of change of the gravitational-wave chirp
frequency. Note that within the simple Newtonian approximation presented here
(at the leading order of the post-Newtonian expansion) the product $h\tau$ is
independent of the components' masses \citep{KrolakS1987}. The simplified
analysis presented above does not take into account the full post-Newtonian
waveform, polarization information, network of detectors analysis etc., but is
intended to demonstrate that binary systems are indeed truly extraordinary
''standard sirens''. Their observations provide absolute, physical distances
directly, without the need for a calibration or a `distance ladder'. Among
them, binary black hole systems occupy a special position. In the framework of
general relativity, binary black holes waveforms are independent from
astrophysical assumptions about the systems' intrinsic parameters and their
environment. At cosmological scales the distance $r$ has a true meaning of the
{\it luminosity distance}. However, since the vacuum (black-hole) solutions in
general relativity are scale-free, the measurements of their waveforms alone
cannot determine the source's redshift.  The parameters measured by the
detector are related to the rest-frame parameters by the redshift $z$: $f_{GW}
= f^{rf}_{GW}/(1 + z)$, $\tau = \tau^{rf}(1 + z)$, $\mathcal{M} =
\mathcal{M}^{rf}(1 + z)$. Independent measurements of the redshift which would
facilitate the cosmographic studies of the large-scale Universe requires the
collaboration with the electromagnetic observers i.e., the {\it multi-messenger
astronomy}. This may be obtained by assessing the redshift of the galaxy
hosting the binary by detecting the electromagnetic counterpart of the event
\citep{Bloom2009,Singer2016,GW170817MMA}, or by performing statistical study for galaxies'
redshifts correlated with the position of the signal's host galaxy
\citep{MacLeodH2008}.The omnidirectional nature of a solitary detector is
mitigated by simultaneous data analysis from at least three ground-based
detectors in order to perform the triangulation of the source position, hence
the crucial need for the LIGO-Virgo collaboration, which will be in the future
enhanced by the LIGO-India detector, and the KAGRA detector in Japan
\citep{KAGRA2013}. 
The above principles were validated by the observations 
of a binary black-hole merger with three detector network \citep{GW170814}, 
resulting in improved distance and sky localization.

Compact binaries containing matter are perfect sources to
produce the multi-messenger events. Neutron-star binaries are now firmly
established as sources of short gamma-ray bursts (idea originally proposed
in \citealt{Paczynski1986}), thanks to simultaneous gravitational and electromagnetic-wave
observations from the same event (\citealt{GW170817MMA,GW170817GRB} and references 
therein). 
Although the complete waveform which
includes the merger requires the knowledge of the material properties of the
components (the presently poorly-known dense-matter equation of state physics),
the chirp waveforms are understood well enough to facilitate a firm detection
and a distance measurement. Binaries involving neutron stars are the ideal ''standard sirens'' 
insofar as they naturally provide both loud gravitational-wave and bright
electromagnetic emission. Short gamma-ray bursts occur frequently within the
reach of ground-based detectors, at redshifts $z<0.2$. \citet{Nissanke2013} show 
that observing a population of the order of 10 of gravitational-wave events related 
to short gamma-ray bursts would allow to measure the Hubble constant with 5\%
precision using a network of detectors that includes advanced LIGO and Virgo 
(30 beamed events could constrain the Hubble constant to better than 1\%). 
In both cases of double black-hole binaries and those involving neutron stars, 
rapid electromagnetic counterpart observations and precise catalogs of galaxies 
are needed \citep{AbbottLRR2016,Singer2016}.  

The main source of error in the distance measurement is the limited sensitivity of
the detectors (finite signal-to-noise), which translates to a limited
knowledge of the source's direction and orientation (see e.g.,\citealt{Nissanke2010} 
for a short gamma-ray burst related study). This may be
improved with the measurements of gravitational-wave polarizations with a network 
of three or more detectors. Second limiting factor is the detectors'
calibration uncertainties. 
Recent direct detectors of binary black-hole mergers by two
Advanced LIGO detectors established the distances with rather large error bars
mostly due to these factors (see e.g., \citealt{AbbottProp2016}); the
order-of-magnitude improvement in sky localization was evident for 
the first triple-detection with the global network of
Advanced LIGO and Advanced Virgo detectors \citep{GW170814}. Presence of a
third detector in the network proved to be crucial for the precise sky
localization of a binary neutron-star merger \citep{GW170817MMA}, allowing for
rapid electromagnetic follow-up.  
For redshifts larger than $z=1$ weak gravitational lensing will contribute to the distortion of the
luminosity distance measurements at the order of 10\%
\citep{BartelmannS2001,Dalal2006}. For a detailed discussion of limiting factors 
in the case of a network of detectors see  \citet{Schutz2011}
and references therein.

To conclude, observations of gravitational-waves from cosmological distances
with current and planned detectors (e.g., spaceborne LISA, sensitive to low
frequencies corresponding to chirping super-massive black hole binaries,
\citealt{Danzmann1996}, or a third-generation cryogenic underground Einstein
Telescope, with an extended low frequency range compared to the Advanced LIGO and Advanced Virgo,
\citealt{ET2011}) promise a wealth of new astrophysical information, 
as demonstrated by the first multi-messenger 
gravitational-wave event \citep{GW170817MMA,GW170817GRB}.
Future detectors will reach cosmological distances and redshifts of a few, being
sensitive to practically all the chirping binaries in their sensitivity band in
the Universe and providing precise distance measurements (see e.g.,
\citealt{LangH2008}). In addition to precisely measuring the Hubble constant,
cosmological observations would help determine the distances to galaxies, thus
contributing to building the standard 'distance ladder' (calibrating
electromagnetic standard candles), establish the distribution of galaxies and
voids, characterize the evolution of the dark energy and mass density of the
Universe, mass distribution through the gravitational lensing, as well as the
chemical evolution effects i.e., establishing the onset of star formation 
\citep{KrolakS1987,SathyaprakashS2009}. The truly multi-messenger era of astronomy 
is just beginning.

\section{Conclusions}

The coming years will bring new large data sets at various wavelengths of the electromagnetic spectrum, accompanied by a new window due to the discovery of gravitational waves. Dedicated projects and generally oriented space and ground-based observatories will allow to make the distance measurements with current well developed methods more precise, and the new emerging methods have a chance to reach maturity. Large data sets will push the statistical errors down significantly, so the possible systematic biases will become more important. Improvements within a given method will be crucial, but the key tests will lie in cross-check of the results based on independent methods. From this point of view, new emerging methods are as important as the further increase in the already impressive accuracy achieved with the use of SN Ia. 

\begin{acknowledgements}
BCZ was supported by the NCN grant 2015/17/B/ST9/03436 and a Fellowship from the Chinese Academy of Sciences, and acknowledges being a part of the network supported by the COST Action TD1403 {\it Big Data Era in Sky and Earth Observation}.  MB was partially supported by the NCN grants UMO-2014/14/M/ST9/00707, UMO-2013/01/ASPERA/ST9/00001 and the COST Action MP1304 ''NewCompStar''. RFIH was supported by Conselho Nacional de Desenvolvimento Cientifico e Tecnologico (CNPq). SWJ acknowledges support from NASA/Keck JPL RSA 1508337 and Department of Energy award DE-SC0011636. JJ gratefully acknowledges the support of NASA grant GO-14219.003-A.  E.L. is supported by a European Union COFUND/Durham Junior Research Fellowship (under EU grant agreement no. 609412). MDO acknowledges partial support by the PRIN-INAF 2014, 
"Transient Universe: unveiling new types of stellar explosions with 
PESSTO" (P.I.: A. Pastorello). We thank ISSI-BJ for hospitality and an engaging workshop. 
\end{acknowledgements}

\bibliographystyle{spbasic}      
\bibliography{ref_chapter5}   
\end{document}